%% file: main.tex
\documentclass[acmsmall,screen,nonacm]{acmart}
\usepackage[english]{babel}

\usepackage[utf8]{inputenc}
\usepackage{braket}
\usepackage{multirow}
\usepackage{tikz}
\usepackage{quantikz}
\usepackage{proof}
\usepackage{relsize}
\usepackage{mathtools}
\usepackage{mathpartir}
\usepackage{tcolorbox}
\usepackage{subcaption}
\usepackage{listings}
\usepackage{colortbl}
\usepackage{cleveref}
\usepackage{graphicx}
\usepackage{comment}
\usepackage{wrapfig}
\usepackage{textcomp}
\usepackage[ruled]{algorithm2e}
\usepackage{algpseudocode}
\usepackage{xspace}
\usepackage{stmaryrd}
\usepackage{paralist}
\usepackage{enumitem}
\usepackage{mdframed}

\usepackage{grumble}

\usepackage[nounderscore]{syntax}
\usepackage{color}
\usepackage{adjustbox}

\setlist[enumerate]{
    leftmargin=1.5em,   
    label=\arabic*),  
    itemsep=2pt       
}

\AtBeginDocument{%
  }

\setcopyright{acmlicensed}
\copyrightyear{2018}
\acmYear{2018}
\acmDOI{XXXXXXX.XXXXXXX}
\acmConference[Conference acronym 'XX]{Make sure to enter the correct
  conference title from your rights confirmation email}{June 03--05,
  2018}{Woodstock, NY}

\acmISBN{978-1-4503-XXXX-X/2018/06}

\input{space-tricks}
\begin{document}

\newcommand{\projectname}{\textsc{MultiQ}\xspace}


\newcommand{\circuitlevel}{Circuit-level\xspace}
\newcommand{\regionIR}{Region-level\xspace}
\newcommand{\tileIR}{Tile-level\xspace}
\newcommand{\logicalIR}{Logical-level\xspace}
\newcommand{\physicalIR}{Physical-level\xspace}

\newcommand{\qtir}[1]{$\texttt{QTIR}_{#1}$}
\newcommand{\QTIR}{\textsc{QTIR}\xspace}

\newcommand{\grammarRuleDefColor}{YellowOrange}
\newcommand{\grammarTerminalColor}{LimeGreen}
\newcommand{\grammarSymbolColor}{RawSienna}
\newcommand{\grammarDefinedNonterminalColor}{Turquoise}
\newcommand{\grammarTypeColor}{Orchid}

\newcommand{\grammarOp}[1]{\texttt{\textcolor{\grammarRuleDefColor}{#1}}}
\newcommand{\grammarArg}[1]{\texttt{\textcolor{\grammarTerminalColor}{#1}}}
\newcommand{\grammarType}[1]{\texttt{\textcolor{\grammarTypeColor}{#1}}}

\newcommand{\researchquestion}[2]{
\begin{mdframed}[%
frametitle={%
\tikz[baseline=(current bounding box.east),outer sep=0pt]
\node[anchor=east,rectangle,fill=white]{\strut {#1}};
},%
linecolor=black!70,
innertopmargin=-4pt,
innerbottommargin=6pt,
skipabove=1pt,
innerleftmargin=8pt,
innerrightmargin=14pt,
linewidth=1pt,topline=true,%
frametitleaboveskip=\dimexpr-\ht\strutbox\relax
]%
{#2}%
\end{mdframed}%
}

\newcommand{\tinstr}[2]{\mathtt{#1} [#2]}

\title{\projectname: Multi-Programming Neutral Atom Quantum Architectures}

\author{Francisco Romão}
\affiliation{
  \institution{Technical University of Munich}
  \city{Munich}
  \country{Germany}
}
\author{Daniel Vonk}
\affiliation{
  \institution{Technical University of Munich}
  \city{Munich}
  \country{Germany}
}
\author{Emmanuil Giortamis}
\affiliation{
  \institution{Technical University of Munich}
  \city{Munich}
  \country{Germany}
}
\author{Dennis Sprokholt}
\affiliation{
  \institution{Technical University of Munich}
  \city{Munich}
  \country{Germany}
}
\author{Pramod Bhatotia}
\affiliation{
  \institution{Technical University of Munich}
  \city{Munich}
  \country{Germany}
}
\authorsaddresses{}

\maketitle
\newcommand{\ie}[0]{\textit{i.e.,}}
\newcommand{\eg}[0]{\textit{e.g.,}}
\input{sections/abstract}

\input{sections/introduction}

\input{sections/background}
\input{sections/motivation}

\input{sections/overview}

\input{sections/compiler}

\input{sections/controller}
\input{sections/consistency-checker}

\input{sections/evaluation}

\input{sections/related_work}
\input{sections/conclusion}

\newpage

\bibliographystyle{ACM-Reference-Format}
\bibliography{references}

\newpage
\input{sections/appendix}

\end{document}

%% file: space-tricks.tex
\usepackage[subtle,margins=normal]{savetrees}

\usepackage{setspace}

%% file: sections/abstract.tex
\vspace{-5mm}
\myparagraph{Abstract} Neutral atom Quantum Processing Units (QPUs) are emerging as a popular quantum computing technology due to their advantages, including large qubit counts and flexible connectivity. However, a key performance trade-off exists: large circuits suffer significant drops in fidelity, yet small circuits underutilize available hardware and are dominated by initialization latency. These issues result in inefficient hardware utilization and limit overall system performance. To address this challenge, we propose {\em multi-programming on neutral atom QPUs}, i.e., co-executing multiple circuits on the same QPU by logically partitioning the large qubit array, enabling increased resource utilization (amortizing initialization latency across jobs), while preserving result fidelity (by efficient hardware circuit mapping and reducing overall circuit size).

Unfortunately, the state-of-the-art compilers for neutral atom architectures do not support multi-programming. To address this research gap, we propose \projectname, the first system to enable multi-programming on neutral atom QPUs. \projectname addresses three key challenges with a set of key ideas. \textit{(i)} To maximize spatio-temporal hardware utilization, we compile circuits to fit in a \emph{virtual zone layout}, independent from specific hardware. We bundle multiple virtual layouts to fit the available hardware qubits before execution. \textit{(ii)} To maximize throughput, we parallelize the execution of co-located circuits, making a single hardware instruction operate on qubits belonging to different independent circuits. \textit{(iii)} To ensure the parallelization did not erroneously introduce new behaviors, we employ an algorithm that checks whether the bundled circuits are functionally independent.

We implement \projectname as a cross-layer system spanning a compiler, (runtime) controller, and checker. Our compiler produces \textit{virtual zone layouts}, maximizing hardware utilization and circuit performance. \projectname's controller efficiently maps these layouts on the hardware, minimizes execution latency, and resolves concurrent operation conflicts. Finally, \projectname's checker ensures the circuits are bundled correctly.

Our results show a throughput increase from 3.8$\times$ to 12.3$\times$ when multi-programming 4 to 14 circuits, respectively. \projectname maintains individual circuit fidelity to a high extent, from a 1.3\% improvement for four circuits to a minimal loss of 3.5\% for 14 circuits. Overall, \projectname strives for seamless concurrent execution of multiple quantum circuits on a given hardware QPU, thereby increasing throughput and hardware utilization.

%% file: sections/introduction.tex
\section{Introduction}
\label{sec:introduction}

\begin{figure}
    \centering
    \includegraphics[width=\textwidth]{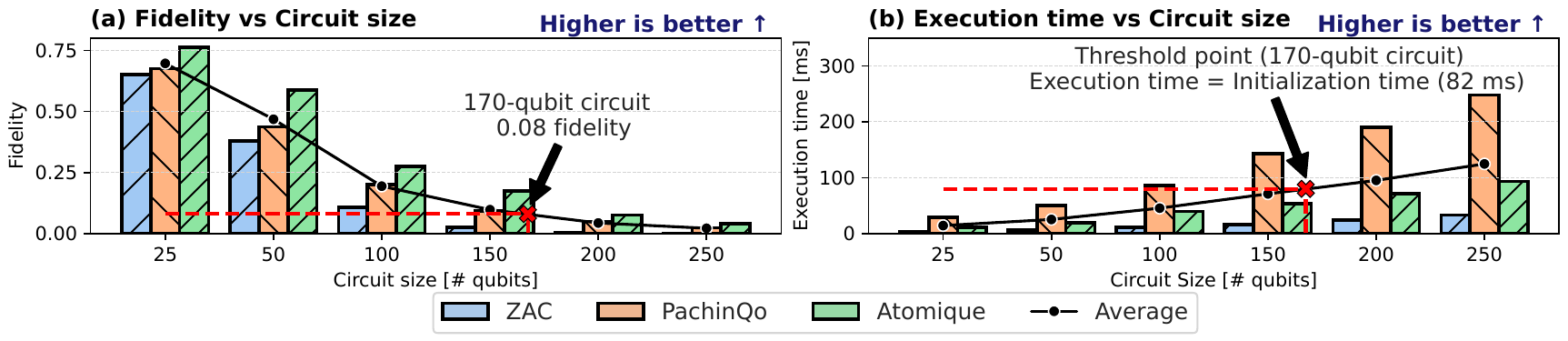}
    \vspace{-17pt}
    \caption{\textbf{(a)} Limitations of neutral atom QPUs evaluated using state-of-the-art NA compilers (ZAC~\cite{lin2024reuseawarecompilationzonedquantum}, PachinQo~\cite{ludmir_modeling_2024} and Atomique~\cite{wang_atomique_2024}).
    {\em (a) Fidelity drops drastically with circuit size, leading to QPU underutilization. (b) Circuit execution time is lower compared to QPU initialization time for circuits up to 170 qubits.}}
    \label{fig:introduction_plots}
    \Description{}
    \label{fig:introduction_background}
    \vspace{-3pt}
\end{figure}

Quantum computing promises significant performance increases for key problems, such as integer factorization and quantum chemistry simulations~\cite{arute2019quantum, peruzzo2014variational}. A variety of physical platforms, including superconducting~\cite{siddiqi2021engineering}, trapped ions~\cite{haffner2008quantum}, and neutral atoms (NA)~\cite{bluvstein2024logical}, aim to realize this potential. Among these, NA Quantum Processing Units (QPUs) are emerging as a leading technology~\cite{saffman2016quantum, ebadi2021quantum}, offering several advantages, including long coherence times~\cite{picken2018entanglement, evered2023high}, flexible connectivity with dynamic trap reconfiguration~\cite{bluvstein2022quantum, beugnon2007two}, native multi-qubit gates~\cite{graham2022multi, bluvstein2024logical} and the scalability to hundreds or even thousands of qubits~\cite{manetsch2024tweezer, singh2022dual}. The NA technology is based on a grid of atoms, such as Cesium or Rubidium, held in space through optical tweezers in a geometric configuration \cite{evered2023high}. Recent NA hardware features distinct zones for different operations, such as an entanglement zone for two-qubit gate applications, a storage zone for idle atoms, and a measurement zone for atom readout \cite{evered2023high}.

Current NA QPUs face two core problems that limit their performance: low fidelity and throughput. We empirically demonstrate these issues, which motivate our research.

\smallskip
\myparagraph{The fidelity problem} Despite their large scale, current NA QPUs suffer from relatively high operation noise, as each execution step (quantum gate) is not perfect, leading to small errors accumulating throughout execution. Fidelity quantifies how close the observed output is to the theoretical ideal, on a scale from $0{-}1$, where $1$ denotes a noiseless result. As the number of qubits in a circuit grows, fidelity drops sharply, making much of a large QPU effectively unusable. Figure~\ref{fig:introduction_plots}~(a) highlights this issue: on a 250-qubit device, with state-of-the-art compilers~\cite{wang_atomique_2024, ludmir2025modeling, lin2024reuseawarecompilationzonedquantum}, on average, estimated fidelity falls below 0.5 for circuits exceeding 50 qubits---just 20\% of the total 250 available qubits.

\smallskip
\myparagraph{The throughput problem} NA QPUs face throughput limitations from two factors: on one side, the fidelity problem restricts the hardware space that can be effectively utilized, and on the other side, NA QPUs have a time-consuming initialization process that creates a relatively high execution latency. Specifically, a NA initialization procedure must run before every circuit execution, starting by loading atoms into a vacuum chamber, imaging them, and sorting them into their correct positions~\cite{blochtrapping2023, schymik2020enhanced}. These tasks incur a latency that can take tens of milliseconds~\cite{Wurtz2023aquila, sheng2021efficient} before the quantum circuit can start to execute. Figure~\ref{fig:introduction_plots}~(b) illustrates this problem: initializing a 250-qubit NA QPU amounts to around 82 ms (blue dotted line)~\cite{Wurtz2023aquila, sheng2021efficient}. In contrast, the actual circuit execution (black solid line) typically takes at most tens of milliseconds and is shorter than the initialization latency time for common circuits up to 170 qubits. As a result, the total runtime is dominated by these QPU initialization overheads, not the computation itself. Figure~\ref{fig:introduction_plots}~(a) shows that a circuit this size would be able to achieve around 0.08 fidelity, producing mostly unusable results.

In summary, while NA QPUs are scaling rapidly, large circuits suffer significant drops in fidelity; yet, small circuits underutilize available hardware and are dominated by initialization latency.

A promising solution to tackle both the fidelity and throughput problems is {\em multi-programming}, where several circuits execute simultaneously on the QPU \cite{das_case_2019, giortamis2025qos}. By executing multiple small circuits concurrently, multi-programming increases overall QPU utilization, as a larger percentage of the QPU's qubits are actively used. This, combined with amortizing high initialization costs across all co-scheduled circuits, significantly improves QPU throughput. Furthermore, by strategically placing these circuits, multi-programming helps reduce execution contention, thereby maintaining the high circuit fidelity inherent to smaller-scale execution. Despite these benefits, state-of-the-art NA compilers, such as ZAC~\cite{lin2024reuseawarecompilationzonedquantum}, PachinQo~\cite{ludmir_modeling_2024}, and Atomique~\cite{wang_atomique_2024}, are only designed to handle single-circuit execution and lack multi-programming optimizations. Realizing multi-programming on NA QPUs requires solving three key challenges:

\begin{enumerate}
    \item \textbf{Maximizing spatio-temporal hardware utilization} -- To maximize QPU throughput, we must co-optimize for both spatial utilization (allocated space) and temporal utilization (active computing time). This creates a complex packing problem, as opting to place more circuits on the same QPU will complicate optimal circuit-runtime matching from a pool of circuits.
    \item \textbf{Maximizing instruction parallelization} -- While serializing circuit execution is a straightforward approach to prevent hardware resource conflicts, it sacrifices parallelism, increasing execution runtime. Parallelizing instructions is ideal for achieving high throughput; however, optimally resolving QPU resource contention is a challenging problem to solve.
    \item \textbf{Preserving functional independence} -- Co-located circuits must yield the same results as they would when executed in isolation. To guarantee this, we must first establish a formal definition of correctness for multi-programmed execution. This is essential for identifying and preventing any resource conflicts that could violate execution independence.
\end{enumerate}

\noindent
We capture those challenges in the main research question of this work:

\researchquestion{Research Question}{\em How can we multi-program NA QPUs, maximizing throughput and minimizing fidelity loss, while ensuring functional independence?}

\vspace{5pt}
\noindent
To address those challenges by introducing \projectname, a compiler-runtime co-design for multi-programming NA QPUs. \projectname achieves high fidelity, utilization, and throughput, ensuring that the final results are identical to those obtained through independent execution.

\smallskip
\myparagraph{Key ideas}
Our key ideas are:

\begin{inparaenum}[\em(1)]
    \item We introduce the novel concept of {\em virtual layouts} to decouple compilation from specific hardware placement. This abstraction uses an efficient balancing formula to independently allocate virtual hardware space to each circuit before finding a physical location. Building on this, we introduce a greedy algorithm that processes these virtual layouts to find near-optimal circuit bundles, simultaneously optimizing both spatial and temporal hardware utilization.

    \item To maximize throughput, we \emph{parallelize} the execution of co-located circuits. Our scheduler analyzes the instruction streams of all active circuits concurrently to identify opportunities for SIMD-like (Single Instruction, Multiple Data) parallelization, where a single hardware instruction can be broadcast to operate simultaneously on qubits belonging to different, independent circuits.
    
    \item We formally define \emph{functional independence}, which mandates that the semantics of a circuit under multi-programming must remain identical to its execution in isolation, even when instructions are shared across circuits. To enforce this, we use a circuit analysis algorithm that leverages ZX-diagrams \cite{coecke_interacting_2011} and ZX-calculus graph optimization techniques. This algorithm performs a scalable, formal verification of semantic equivalence between the isolated and co-executed versions of each circuit, ensuring that no unintended cross-circuit interactions are introduced.
\end{inparaenum}

\smallskip
\myparagraph{\projectname: A compiler-runtime co-design} Our system, \projectname, realizes these key ideas through a compiler-controller co-design, consisting of three main components.
\begin{inparaenum}[\em(i)]
    \item Our \emph{compiler} produces an optimized executable with a corresponding virtual layout for a given circuit, balancing hardware utilization and circuit performance.
    \item Our \emph{controller}, a runtime component, bundles multiple virtual qubit layouts into hardware-fitted bins, balancing temporal and spatial QPU utilization.
    It then schedules independent circuit instructions in a unified executable that minimizes resource contention.
    \item Finally, our \emph{checker} determines whether the multi-programmed executable ensures functional independence of the original components, thus ensuring that bundling did not introduce errors.
\end{inparaenum}

\smallskip
\noindent
We integrate \projectname with existing toolchains, including the Qiskit transpiler for basic circuit optimizations~\cite{qiskit-transpiler} and the ZAC compiler~\cite{lin2024reuseawarecompilationzonedquantum} for solo circuit compilation. Our results, based on 11 standard applications, demonstrate that \projectname delivers a significant throughput improvement, ranging from 3.8$\times$ to 12.3$\times$ when multi-programming 4 to 14 circuits, respectively. Additionally, the system maintains high individual circuit fidelity, with an improvement of 1.3\% achieved by multi-programming four circuits and a minimal loss of 3.5\% for 14 circuits.

\smallskip
\myparagraph{Contributions} \projectname makes the following contributions:
\begin{enumerate}
    \item \textbf{Efficient NA multi-programming} -- \projectname is the first system to efficiently and scalably co-execute multiple circuits on NA QPUs, while preserving the fidelity of individual circuits.
    \item \textbf{Novel virtual zone layout} -- We introduce the concept of a \textit{virtual zone layout}, enabling independent compilation and optimization of multiple quantum circuits, allowing circuit bundling before being assigned to specific hardware.
    \item \textbf{Instruction-parallelization optimizations} -- We present new instruction-parallelization optimizations that enhance circuit fidelity in both solo and multi-programming environments, in comparison to existing compilation methods that are unaware of multi-programming.
    \item \textbf{Functional independence checker for multi-programming} -- We present the first method to systematically check functional independence between multi-programmed quantum circuits, ensuring circuits behave the same in solo and multi-programming environments.
\end{enumerate}

%% file: sections/background.tex
\section{Neutral Atom (NA) Quantum Architectures}
\label{section:background}

\subsection{Quantum Computation}
A quantum computation denotes a quantum circuit acting on $m$ qubits, initialized in the computational state $\ket{0}^{\otimes m}$, where $\otimes$ represents the tensor product of $m$ qubits initialized in $\ket{0}$. A circuit operates on $m$ qubits by applying a sequence of gates, $U = U_L \cdots U_2 U_1$, where each gate $U_i$ corresponds to either a single-qubit or multi-qubit operation. A gate can be any rotation or a linear combination of different rotations on the axes $x,y,z$. These gates transform the initial state to a final state $\ket{\psi} = U\ket{0}^{\otimes m}$. Finally, the circuit ends with a final measurement of the expectation value of an observable $O$, denoted as $\braket{O} = \bra{\psi}O\ket{\psi}$.
While theoretical quantum computing can be realized through various hardware technologies, this paper focuses on neutral atom (NA) technology. In the following sections, we provide more details on the capabilities and limitations of this technology.

\subsection{Neutral Atom (NA) Architectures and Characteristics}

\begin{wrapfigure}[18]{r}{0.5\textwidth}
    \centering
    \vspace{-2pt}
    \includegraphics[width=\linewidth]{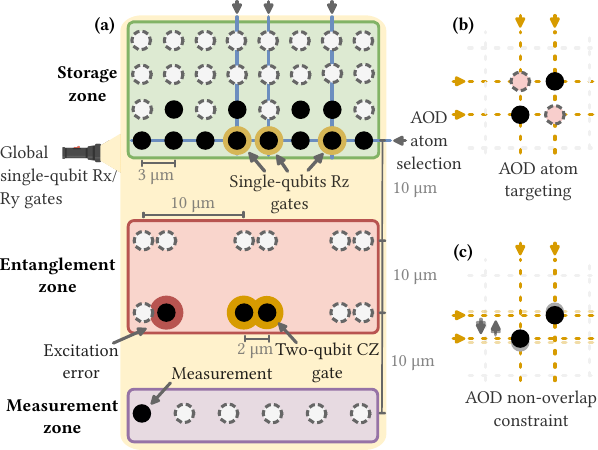}
    \caption{Neutral atoms architecture basics (\S~\ref{section:background}) \textit{Storage, entanglement, and measurement zones distributions and their standard atom and zone spacings. Single-qubit gates and two-qubit gate operations. AOD laser targeting and the non-overlapping constraints.}}
    \label{fig:background}
\end{wrapfigure}
NA quantum architectures utilize arrays of NAs, commonly alkali species such as rubidium, cesium, or strontium, which are excited into high-energy Rydberg states to encode qubits \cite{saffman2010quantum, briegel2000quantum, jaksch2000fast}. Atom arrays are held in place by static spatial light modulators (SLMs). 
This architecture enables multi-qubit gates and supports dynamic qubit rearrangement through acoustic-optical deflectors (AODs), allowing practical all-to-all qubit connectivity. Two-qubit gates are typically realized using an optical beam, also known as an entanglement pulse. To reduce crosstalk and noise on non-interacting atoms, modern architectures divide the system into distinct zones for entanglement, storage, and readout, thereby restricting the entanglement pulse and readout pulses to their respective zones. 

\smallskip
\myparagraph{NA capabilities}
\label{sec:back:types_properties}
NA QPUs offer unique capabilities compared to superconducting ones, including native $\geq 2$ qubit gates \cite{bluvstein2022quantum, beugnon2007two, graham2022multi}, longer decoherence times \cite{picken2018entanglement, evered2023high}, and reconfigurable qubit layouts \cite{wintersperger_neutral_2023, bluvstein2022quantum, bluvstein2024logical}. Moreover, NA QPUs show promising scalability, with current commercial QPUs already having hundreds of qubits \cite{Wurtz2023aquila, singh2022dual}, and near-term QPUs expected to reach the thousands \cite{manetsch2024tweezer}. However, operations in NA QPUs are relatively slow ($\mu s-ms$ timescales), limiting usable circuit depth before decoherence becomes significant \cite{Wurtz2023aquila}.

\smallskip
\myparagraph{Monolithic vs. zoned layouts}
The hardware of common NA QPUs can be set up using either \textit{monolithic layouts}, where all operations share a single zone, or \textit{zoned layouts}, where atom arrays are physically separated into three different zones: \textit{storage}, \textit{entanglement}, and \textit{measurement}, as shown in Figure~\ref{fig:background}~(a). Zoned architectures, which \projectname uses, are increasingly preferred for improving fidelity by isolating idle qubits and allowing mid-circuit measurements \cite{stade_abstract_2024, ludmir_modeling_2024}.


\myparagraph{Gate operations}
\label{sec:gate_operations}
NA QPUs natively support single- and two-qubit gates. Two-qubit gates are based on the Rydberg blockade mechanism, as illustrated in Figure~\ref{fig:background} (a): two atoms inside each other's blockade radius ($2-4~\mu m$) cannot both be excited to Rydberg levels, enabling a controlled-Z entanglement gate~\cite{schmid2024computational, bluvstein2022quantum}. Single-qubit gates can be applied locally or globally. Local gates are limited to rotations around the Z-axis, which can be applied to multiple atoms selected using AOD lasers, while adhering to the AOD targeting rules~\cite{evered2023high, graham2022multi, bluvstein2024logical}. As illustrated in Figure~\ref{fig:background} (b), diagonal atoms cannot be selected without selecting the atoms on the opposing diagonal, and Figure~\ref{fig:background} (c), the top row cannot cross the bottom row. In this work, we focus on single- and two-qubit gates.

\smallskip
\myparagraph{Laser and trap system}
NA arrays use two optical trap systems \cite{beugnon2007two}. Spatial light modulators (SLM) create arbitrary 2D trap patterns to statically hold atom arrays, while acousto-optic deflectors (AOD) enable dynamic repositioning of qubits at runtime \cite{ebadi2021quantum, bluvstein2024logical}. Figure \ref{fig:background}~(a) shows grids of SLM traps (white and black circles) and AOD lasers. A single AOD laser can manipulate multiple rows and columns of atoms in parallel \cite{schmid_computational_2024, bluvstein_quantum_2022, evered_high-fidelity_2023}. AOD lasers are subject to constraints such as active lasers cannot cross over each other, or diagonally targetting can select unwanted atoms \cite{bluvstein2022quantum, Wurtz2023aquila, tan2024compilingquantum}.


\smallskip
\myparagraph{Initialization procedure}
\label{sec:background:initialization}
Initialization in NA QPUs contributes significantly to the overall runtime, as we show in Figure~\ref{fig:introduction_plots}~(b). The process begins by loading atoms into a vacuum cell in which approximately 50\% of the SLM trapping sites will be filled during this initial loading phase~\cite{wang_accelerating_2023, schymik2020enhanced}. The atom array is then imaged to determine the coordinates of the scattered atoms. Then, the sorting algorithm generates a set of arrangement instructions to build the desired atom layout~\cite{wang_accelerating_2023}, and finally, a second image is taken to verify the correct construction of the grid. If discrepancies are found, a new sorting cycle is initiated until all atoms are in their designated locations~\cite{evered_high-fidelity_2023, Wurtz2023aquila, bluvstein_quantum_2022}. Notably, initialization has a constant cost that only depends on the atom array dimensions; it is independent of the size of the circuits that will be executed.

\smallskip
\myparagraph{Qubit movement}
Qubit shuttling enables the transportation of atoms using mobile optical tweezers, effectively achieving near-perfect fidelity when performed below a speed limit \cite{bluvstein2024logical, tan2024compilingquantum}. However, shuttling operations must avoid collision scenarios where two AOD lasers get within a safe distance of each other, increasing the risk of atom loss \cite{bluvstein2022quantum, Wurtz2023aquila}. Atoms can be transferred between SLM and AOD traps with ${\sim}99.9\%$ fidelity, enabling complete dynamic reconfiguration \cite{beugnon2007two, tan2024compilingquantum}.

\begin{figure}[t]
    \centering
    \includegraphics[width=0.9\linewidth]{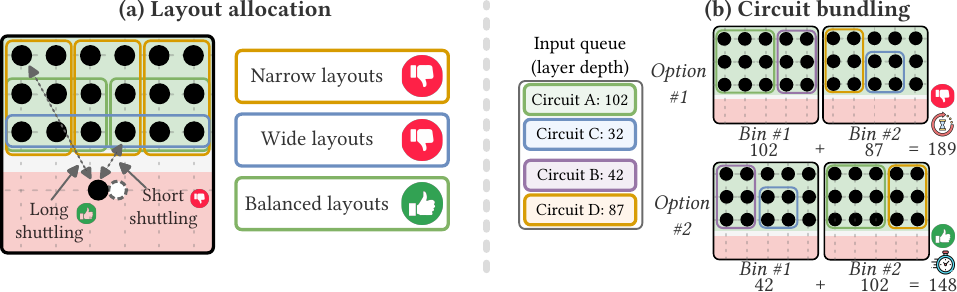}
    \vspace{-2pt}
    \caption{\textbf{(a)} Tradeoff between QPU utilization and circuit shuttling time. \textit{Narrow layouts (orange) fully utilize the QPU but incur long shuttling operations. Wider layouts (red) minimize shuttling times but incur low utilization. Balanced layout (green).} \textbf{(b)} Circuit bundling. \textit{Bundling circuits from the input queue (top) into execution bins (bottom) involves finding a solution that maximizes both spatial and temporal QPU utilization. Here, \textit{Option \#2} reduces the total execution runtime.}}
    \Description{}
    \label{fig:motivation:circuit_distribution}
\end{figure}

%% file: sections/motivation.tex
\section{Motivation} 
\label{sec:motivation}
\projectname mitigates the problem of QPU underutilization and low throughput by introducing multi-programming to the NA technology. Multi-programming increases throughput by co-scheduling multiple circuits onto the same grid, allowing them to execute in parallel without incurring repeated initialization costs. Furthermore, it increases grid utilization by co-scheduling multiple circuits that would independently underutilize the QPU's available qubits.

\subsection{Problem Statement}
\label{sec:prob_statement}
\projectname answers the question proposed in Section \ref{sec:introduction}: {\em How can we multi-program NA QPUs, maximizing throughput and minimizing fidelity loss, while ensuring functional independence?} Intuitively, \emph{throughput} corresponds to the average number of circuits executed per time unit, while \emph{fidelity} captures the closeness of the observed result to the theoretical ideal. We formally define both below. 

\smallskip
\myparagraph{Throughput}
We simultaneously execute multiple circuits tiles in a \emph{bin} $B_j = \{c_{j1}, \dots, c_{jn}\}$. Each circuit is compiled into a tile $c$ with width $w(c) \in \mathbb{Z}_{>0}$ and execution time $t(c) {>} 0$. The tiles in bin $B_j$ can execute simultaneously if they fit in the total QPU width, ${\sum_{c{\in}B_j}}, {w(c)} < W_{QPU}$. Total QPU width is computed as: $W_\text{QPU} = R \cdot W$, where $W$ is the physical width of the hardware space, and $R \in [0,1]$ number of storage rows. The wall-time of executing the bundled circuit $B_j$ is thus $T(B_j) = t_\text{init} + \max_{c \in B_j} t(c)$. 



We can then define the throughput as the number of circuits executed per unit time. If we had executed only a single circuit $c$, our throughput would simply be $\frac{1}{t_\text{init}~+~t(c)}$. However, when scheduling $N$ bins, each with multiple circuit tiles that can co-execute, we can define throughput for that entire set:

\vspace{2pt}
\begin{definition}{Throughput}{def:throughput}
Given $N$ bins $B_j$ (for $1 {\leq} j {\leq} N$), the throughput is calculated as:
\[
\tau = \frac{\sum_{j=1}^N \left| B_j \right|}{\sum_{j=1}^{N} T(B_j)}
\]
\end{definition}
\vspace{2pt}

\noindent

\smallskip
\myparagraph{Fidelity}
Quantum fidelity measures the closeness of a noisy quantum state to the desired ideal target state, expressed as a value between 0 and 1. 
When the ideal state is $\ket{\psi_{ideal}}$ and the noisy state is $\ket{\psi_{noisy}}$, then the fidelity $F$ is defined as:
$
F(\ket{\psi_{ideal}}, \ket{\psi_{noisy}}) = |\braket{\psi_{ideal}|\psi_{noisy}|\psi_{ideal}}|^2
$.
%
%
In practice, especially for large circuits, computing the ideal state $\ket{\psi_{ideal}}$ is not feasible. Therefore, state-of-the-art compilers often estimate overall circuit fidelity based on the known error rates of individual quantum operations and decoherence \cite{lin2024reuseawarecompilationzonedquantum, ludmir2025modeling, wang_atomique_2024}. The general approach to estimate fidelity is: For each qubit $i \in [0{..}N[$, track all applied gates ${g^{(i)}_1, ..., g^{(i)}_n}$; each operation has an associated operation fidelity $f_{g_k}$, and each qubit experiences some decoherence $d^{(i)}(t) = 1 - e^{-t/T2_i}$, where $t_i$ is the idle time of qubit and $T_2$ is the dephasing time.

\vspace{2pt}
\begin{definition}{Estimated Total Fidelity}{def:estfidelity}
We \emph{estimate} the total fidelity for a circuit with $N$ qubits as follows
\vspace{-5pt}
\[
F_{total} \approx \prod_{i=0}^N\left(\prod_{k=0}^{n_i} f_{g_k} \cdot \exp\left(-\frac{t_i}{T_{2_i}}\right)\right) \text{,}
\]
where each qubit $i \in [0{..}N[$ is applied to $n_i$ gates.
\end{definition}
\vspace{2pt}

\noindent
Maximizing total fidelity thus requires minimizing the number of gates per qubit, while prioritizing the ones with the lowest error rates.

\smallskip
\noindent
We must consider both throughput and fidelity when multi-programming circuits on a QPU, which we include in our problem statement as:

\noindent
\begin{minipage}{\textwidth}
\noindent
\begin{center}
\researchquestion{Problem Statement}{
When multi-programming circuits, \projectname tries to simultaneously maximize throughput 
$\tau$ (\Cref{def:throughput}) and preserve the total fidelity $F_{total}$ (\Cref{def:estfidelity}) of the original circuits.
}
\end{center}    
\end{minipage}


\subsection{Design Challenges and Key Ideas}
To address our problem statement, our design builds upon several key ideas, each of which solves a technical challenge. 

\begin{wrapfigure}[17]{r}{0.543\textwidth}
    \centering
    \vspace{-4pt}
    \includegraphics[width=\linewidth]{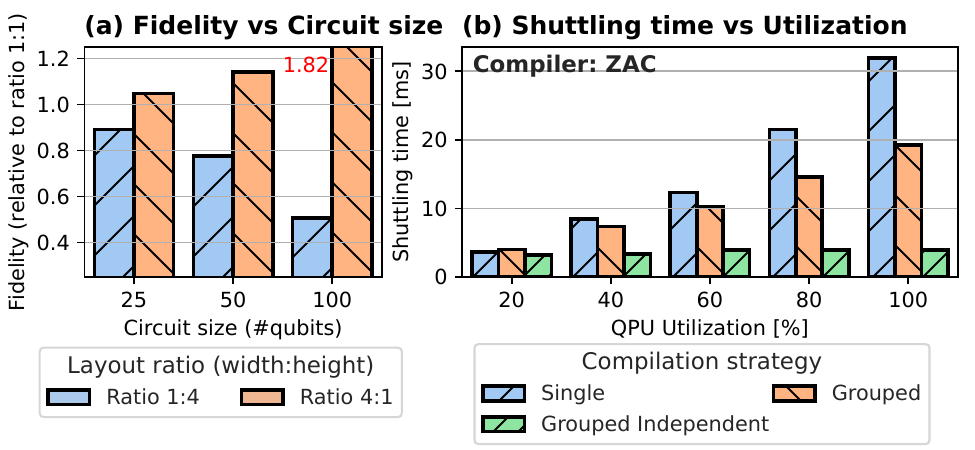}
    \vspace{-16pt}
    \caption{\textbf{(a)} Relative fidelity of two layouts compared to the square layout (ratio 1:1), with increasing circuit size. Narrow layouts (blue bars, 1:4 ratio) achieve lower fidelity than the square ones, while wide layouts (orange bars, 4:1 ratio) achieve higher. \textbf{(b)} Total shuttling time with increasing QPU utilization for ZAC \cite{lin2024reuseawarecompilationzonedquantum}, executing circuit sequentially (single), circuits merged in parallel (grouped), and concurrently and independently (grouped independent)}
    \label{fig:preeval}
\end{wrapfigure}

\noindent
\vspace{-15pt}
\subsubsection{Hardware Utilization}
\label{sec:motivation:resource_allocation}
To simultaneously execute multiple quantum circuits on a QPU, each circuit must first be mapped to a region of hardware space. For a given circuit, this region is referred to as its \textit{layout}, which affects both throughput and fidelity. As Figure \ref{fig:motivation:circuit_distribution}(a) shows, narrow layouts (orange) use space better, allowing to fit more circuits, but each circuit runs slower due to long shuttling paths. In contrast, wider layouts (blue) reduce shuttling, benefiting single-circuit performance, but fit fewer circuits on the QPU. Instead, balanced layouts (green) offer a middle ground. Figure \ref{fig:preeval} (a) shows the relative fidelity of narrow (1:4) vs wide layouts (4:1) in relation to a square layout (1:1). Additionally, when bundling layouts on a QPU, they must be effectively bundled to best fit the available QPU space, maximizing spatial utilization, and match runtime-wise to maximize temporal utilization, as exemplified in Figure \ref{fig:motivation:circuit_distribution}.

\noindent
In summary, maximizing throughput and fidelity requires: (i) balancing single-circuit performance against a smaller layout footprint, (ii) ensuring each bundle best utilizes the QPU space and time resources. We thus phrase this challenge as:



\begin{center}
\noindent\fbox{\parbox{0.98\columnwidth}{
\myparagraph{Challenge \#1} \textit{How can we efficiently allocate space regions for multiple circuits and bundle them, while maximizing throughput and preserving fidelity?}}}
\end{center}

\noindent
We analyze and address this challenge at two levels:
\begin{inparaenum}[\em(i)]
    \item ~First, each circuit is computed a \textit{virtual layout} that balances a smaller layout footprint, which allows more circuits to fit on the QPU space, and high circuit performance.
    \item ~Second, given a large collection of circuits, we must bin them in such a way that each bin will fit on the QPU, while also maximizing throughput across all bins.
\end{inparaenum}

\smallskip
\myparagraph{Fidelity and throughput of multi-programmed circuits}
At the lowest level, we aim to minimize the layout's footprint while maximizing  circuit performance (low runtime and high fidelity) for multiple circuits. However, fidelity is difficult to compute exactly (as seen in Definition \autoref{def:estfidelity}), especially when considering multiple layouts. Instead, we make the following observation:

\begin{center}
\noindent\fbox{\parbox{0.98\columnwidth}{

\smallskip
\myparagraph{Key idea \#1A}
Fidelity is primarily influenced by the width of a circuit, while throughput is primarily affected by the spatial utilization of QPUs. We can thus use spatial utilization as a proxy variable for throughput, which can be checked more efficiently; we can \emph{estimate} fidelity from the layout width. We use these insights to produce a \emph{virtual layout} of the circuit that balances between hardware utilization and fidelity, independent from the QPU hardware.
}}
\end{center}

\noindent
We use this key observation in practice by heuristically navigating the space characterized by circuit width and spatial utilization. Given a layout $\ell$, with width $\ell_w$, we define the estimated fidelity as $P_f(c, \ell_w)$ (from Definition \ref{def:estfidelity}), and spatial utilization as $\rho_S(\ell_w) = \ell_w/W_{\text{QPU}}$, where $W_{\text{QPU}}$ denotes the total QPU width. We are thus interested in the layout $\ell$ that maximizes both, which we denote as:
\[
w_\text{opt}(c) = \arg\max_{\ell} \left[ \alpha \cdot P_f(c, \ell_w) + (1 - \alpha) \cdot \rho_S(\ell_w) \right]\text{,}
\]
\noindent
where $\alpha \in [0,1]$ is a weighting parameter controlling the trade-off between fidelity and throughput.

\smallskip
\myparagraph{QPU utilization across all multi-circuits}
At the global level, we are interested in the optimal set of bins for a given collection of circuits.
In a multi-tenant quantum cloud environment, each QPU receives more circuits than it can execute concurrently, requiring us to partition the input queue into temporally separated bundles \cite{giortamis2025qos, liu2021qucloud}. Each bundle must fit within the QPU’s spatial constraints, and its execution time is dictated by the longest-running circuit within it. Consequently, the bundling strategy directly impacts both total and per-circuit latency. Figure~\ref{fig:motivation:circuit_distribution}~(b) shows this effect: the naive FIFO bundling (Option \#1) leads to significantly higher total runtime compared to a latency-aware alternative (Option \#2), despite both achieving identical spatial utilization.

Unlike the layout selection above, we must now consider and reduce the \emph{total} time needed to execute all circuits.
In particular, we now consider the spatial utilization $\rho_S$ and temporal utilization $\rho_T$ of an a bin---instead of single circuit, like before---which are defined as:
\[
\rho_S(B_j) = \frac{\sum_{c \in B_j} w(c)}{W_\text{QPU}} \in (0,1]
\qquad\qquad
\rho_T(B_j) = \frac{\sum_{c \in B_j} t(c') - t(c)}{\left| B_j \right| \cdot t(c')} \quad \text{where} \ c' = \arg\ \max_{c \in B_j}\ t(c)
\]
\noindent
The goal is to compute the maximum of the weighted sum of both utilizations: $\rho(B) = \alpha \cdot \rho_T(B) + (1-\alpha) \cdot \rho_S(B)$, where $\alpha$ is again a tunable weight parameter. The challenge lies in finding an optimal bundle of circuits $B_\text{opt}$ that maximizes utilization while fitting in the QPU area $W_\text{QPU}$:
\[
B_\text{opt} = \arg\ \max_B\ \rho(B) : \sum_{c \in B} w(c) < W_\text{QPU}
\]
The difficulty again lies in decreasing the computational complexity of exploring large sets, in this case, all possible bundle combinations, for which our key idea is:

\begin{center}
\noindent\fbox{\parbox{0.98\columnwidth}{

\smallskip
\myparagraph{Key idea \#1B}
\projectname uses a simulated annealing algorithm to efficiently search the solution space of hardware-fitting circuit bundles, quickly converging on a near-optimal grouping that maximizes both spatial and temporal QPU utilization. 
}}
\end{center}

\subsubsection{Parallel Execution}
\label{sec:motivation:parallelization_challenge}
Maximizing instruction parallelism between the multi-programmed circuits is essential to avoid trivial instruction sequentialization, which would lead to long execution times. 
This is challenging due to the inherent NA hardware constraints on simultaneous single-qubit gates, entanglement pulses, and the concurrent movement of multiple atoms (\S~\ref{section:background}). 
Figure \ref{fig:background} (b) shows that AOD lasers target all the atoms in the intersections of the horizontal and vertical lasers, which can lead to unintentional atom targeting (pink dotted circles).
Moreover, Figure \ref{fig:background} (c) shows the AOD overlapping constraints, where they require a minimal distance to prevent frequency interference.
Last, as explained in \S~\ref{sec:gate_operations}, single-qubit rotation gates can only be applied in parallel row-wise and on rotations around the same axis. We verify this experimentally in Figure \ref{fig:preeval}~(b), where we plot shuttling time (in ms) with increasing QPU utilization.

With ideal parallelization, multi-programmed circuits $c$ in bin $B$ would execute fully in parallel. Then instructions of the bundled circuit $I(c_j) = \{i_0, i_1,\dots, i_{n_j}\}$ execute in the same duration as the longest independent circuit, represented as:
\[
S(B) \leq \left|I(c_\text{longest})\right| \text{~where~} c_\text{longest} = \arg\ \max_{c \in B}\ \left|I(c)\right|
\]
\vspace{-13pt}

\noindent
where $S(B)$ is the instruction schedule assigning each instruction $i \in \cup_{c \in B}\ I(c)$ a start time $S(i)$.
We aim to determine the optimal schedule $S_\text{opt}$ that executes all co-scheduled circuits in $B$ in the shortest possible time.
The schedule must respect instruction dependencies and hardware constraints (\eg{} laser conflicts, row-wise single-qubit rotation, etc); formally, that goal is:

\vspace{-8pt}
\[
S_\text{opt} = \arg\ \min_S\ [\arg\ \max_i\ (S_\text{end}(i))] : i \in \cup_{c \in B} \ I(c) \text{~,}
\]
\vspace{-13pt}

\noindent
where "$\arg \max_i S_\text{end}(i)$" corresponds to the finishing time of the last instruction in schedule $S$.

\smallskip
\begin{center}
\noindent\fbox{\parbox{0.98\columnwidth}{
\myparagraph{Challenge \#2} \textit{How can we efficiently parallelize instructions in a multi-programming environment to execute all co-scheduled circuits in the least amount of time?}

\smallskip
\myparagraph{Key idea \#2} \projectname approaches this NP-hard problem in a greedy manner by producing a dependency and constraint graph, from which we can extract the largest set of executable instructions.
}}
\end{center}

\subsubsection{Correctness}
\label{sec:correctness}
Multi-programming performance requires maximizing parallelism by resolving hardware constraint conflicts (\S~\ref{sec:motivation:parallelization_challenge}). However, such transformations risk altering the program's semantics or introducing unintended interference between co-executing programs. Unlike classical compilation, where correctness is typically preserved through well-defined static rules, quantum multi-programming must account for entanglement, gate non-commutativity, and shared physical resources. Ensuring correctness in this setting demands new abstractions and safeguards that reason about inter-program interactions at compile time.

Formally, correctness requires that each circuit $c_j^{\text{single}} \in B$ preserves its functional behavior under the multi-programmed execution $E_B$. Each original circuit \smash{$c_j^{\text{single}}$} consists of an ordered sequence of gates $G_j = \{ g_{j_0}, g_{j_1}, \dots, g_{j_{m_j}} \}$ acting on a its local qubits $Q_k$. Its overall unitary transformation is:
\vspace{-8pt}
\[
U_j^{\text{single}} = \prod_{i=m_j}^{0} U(g_{j_i})
\]
\vspace{-10pt}

\noindent
A multi-programmed executable $E_B$ is defined as a global gate sequence $G_B = \{g_{B_0}, g_{B_1}, ..., g_{B_M}\}$ operating on the union of qubits $Q = \bigcup_j Q_j: \forall j \in B$.
The multi-programmed executable is defined as $U^\text{multi} = \prod_{i=M}^{1} U(g_{B_i})$, from which we derive $U_k^\text{multi}$, denoting its restriction to the qubits in $Q_k$.

%

\vspace{2pt}
\begin{definition}{Functional Independence}{def:independence}
The individual circuits in a multi-programmed circuit are functionally independent when transformations only observably affect those qubits assigned to circuit $C_k$.
%
Functional independence of circuit $k$ holds if there exists a global phase $\phi \in [-\pi, \pi]$ such that the unitary operator $U_k^\text{single}$ satisfies:

\vspace{-6pt}

$$
U_k^\text{multi}\cdot (U_k^\text{single})^{\dagger} = e^{i\phi}I_k
$$

\noindent
where $I_k$ is the identity operation on the Hilbert space of qubits $Q_k$ and $U^{\dagger}$ is the conjugate transpose (adjoint) of $U$.
\end{definition}
\vspace{2pt}

\noindent
Intuitively, while the unitary transformations are defined over the global qubit state $Q$, only those transformations relevant for each individual circuit should interact with its assigned qubits.
We explicitly state the resulting challenge and our key idea addressing it as:

\begin{center}
\noindent\fbox{\parbox{0.98\columnwidth}{
\myparagraph{Challenge \#3} \textit{How can we ensure correctness in optimized circuit multi-programming by guaranteeing functional independence of the individual circuits and their multi-programmed version?}

\smallskip
\myparagraph{Key idea \#3} \projectname ensures correctness by taking advantage of the quantum reversibility property as well as the ZX-calculus circuit processing capabilities. Functional independence is checked by simplifying, through ZX-calculus \cite{chen_automata-based_2023}, a concatenated circuit composed of the ZX-diagrams of the original and the multi-programmed versions of a circuit. An empty global phased circuit ensures functional equivalence.}}
\end{center}

%% file: sections/overview.tex
\begin{figure*}[t]
    \centering
    \includegraphics[width=\textwidth]{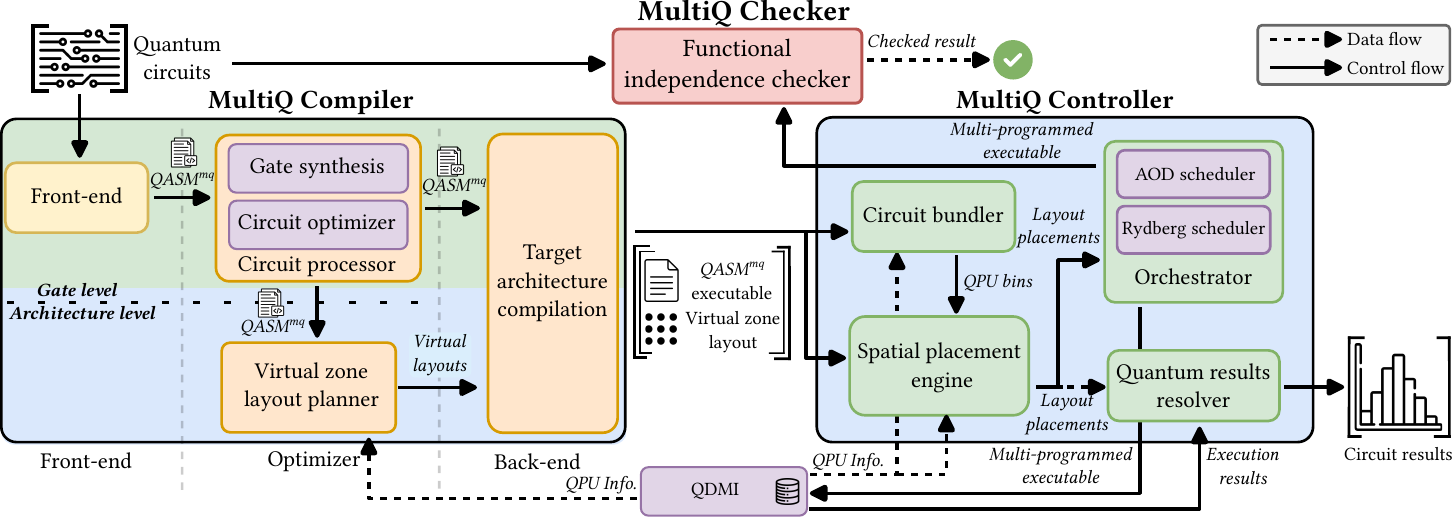}
    \caption{Overview of \projectname (\S~\ref{sec:overview}). \textit{\projectname is a co-designed compiler-controller system. The compiler (\S~\ref{sec:compiler}) optimizes and compiles quantum circuits. The controller (\S~\ref{sec:controller}) then maps and efficiently multiprograms them on the hardware. A functional independence checker (\S~\ref{sec:checker}) verifies that the instructions maintain circuit functionality.}}
    \Description{}
    \label{fig:overview}
\end{figure*}

\section{\projectname Overview}
\label{sec:overview}

We propose \projectname, which addresses the challenges of efficiently executing multiple circuits on a single QPU while preserving fidelity, reducing latency, and increasing throughput.
\projectname consists of three main components: the compiler, the controller, and the checker, as illustrated in Figure \ref{fig:overview}. Next, we explain the function of each main component and how it realizes each key idea.


\smallskip
\myparagraph{\projectname compiler (\S \ref{sec:compiler})}
The compiler starts by independently generating a \textit{virtual zone layout} for each incoming circuit, balancing circuit fidelity and layout footprint, where larger footprints result in lower spatial utilization (\smash{\underline{key idea \#1A}}). Afterwards, the circuit can then be target compiled to the respective virtual layout. Our compiler operates at both gate-level (top, green) and architecture-level (bottom, blue) abstractions. We optimize circuits at the gate level in our QASM dialect (\qasm), while concurrently generating architecture-level virtual layouts.



\myparagraph{\projectname (runtime) controller (\S \ref{sec:controller})}
The controller starts by bundling circuits and their respective virtual layouts into execution bundles using a greedy algorithm that optimizes the spatial and temporal utilization of hardware resources (\smash{\underline{key idea \#1B}}). Secondly, with a set of formed bundles, the controller determines the near-optimal placement of the circuits in a bin, aiming to minimize instruction contention. Finally, the \textit{Orchestrator} schedules hardware resources, producing a non-conflicting multi-programmed executable that executes all the bundled circuits simultaneously (\smash{\underline{key idea \#2}}). Finally, the \textit{Quantum results resolver} maps the results back to their original circuits, delivering separated outcomes to users.

\smallskip
\myparagraph{\projectname checker (\S \ref{sec:checker})}
The functional independence checker ensures the input is semantically equivalent to its embedding in the multi-programmed circuit. This check is necessary to ensure that compiler transformations do not inadvertently alter the behavior of the multi-programmed executable.
We check this by reversing the multi-programmed executable, concatenating it with the solo input circuit, and iteratively eliminating canceling gates (\smash{\underline{key idea \#3}}). We ensure functional independence from the other circuits if the result is an identity circuit.

%% file: sections/compiler.tex
\section{\projectname \compiler}
\label{sec:compiler}

We now explain our \projectname compiler in greater detail. At a high level, it optimizes individual circuits and generates a virtual QPU layout that strikes a balance between circuit fidelity and QPU utilization.

\subsection{\qasm: \projectname Intermediate Representation}
\label{sec:front_end}

The \projectname system uses a front-end to translate quantum circuit descriptions from libraries like Qiskit \cite{ibmTranspilerlatest} or Cirq \cite{quantumaiCirqGoogle} into our intermediate representation called \qasm, which extends the widely-used OpenQASM standard~\cite{openqasm2}. \qasm leverages OpenQASM's annotation features to add NA-specific instructions. Each annotation provides a NA-specific execution of the following hardware-agnostic OpenQASM statement. For example, \qasm extensions include \texttt{@init}, which distributes the atom locations, and \texttt{@move}, which moves an AOD row or column by an offset.

\input{tables/qasm_table}

Table \ref{tab:openqasm} shows the detailed annotations for the \qasm extensions available. \texttt{@init} sets up atom locations on the SLM trap grid, while \texttt{@move} moves one or more atoms; SLM-to-AOD transfers at the starting locations and vice versa at the end locations are implicit in this operation. The \texttt{@u3} operation performs qubit-ID targeted qubit rotations, which can later be optimized to be performed row-by-row or globally. Finally, \texttt{@rydberg} applies a controlled-Z (CZ) gate between qubits within the Rydberg interaction range. Figure \ref{fig:grammar}, in the Appendix \ref{sec:appendix}, formalizes the \qasm grammar in EBNF format.




\subsection{Virtual Zone Layout and Planning}
\label{sec:planner}
\begin{figure}
    \centering
    \begin{subfigure}[c]{0.46\textwidth}
        \includegraphics[width=\textwidth]{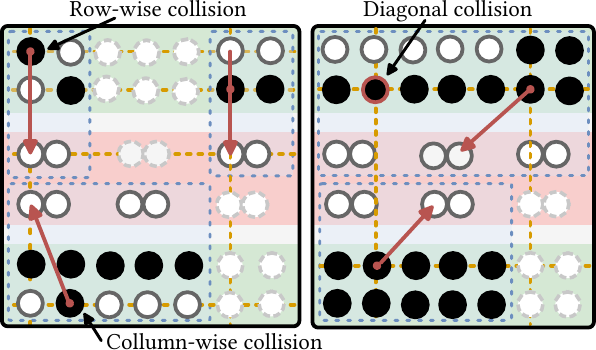}
        \vspace{-0.35cm}
        \caption{Examples of movement collisions (\S~\ref{sec:controller})}
        \label{fig:movements}
    \end{subfigure}
    \begin{subfigure}[c]{0.53\textwidth}
        \includegraphics[width=\textwidth]{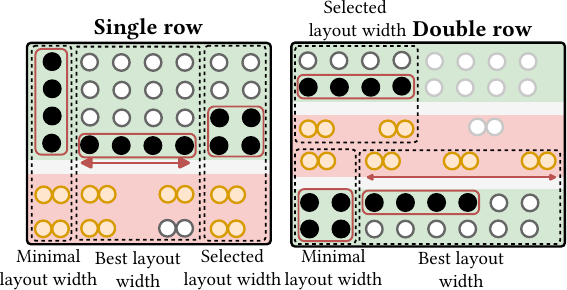}
        \vspace{-0.4cm}
        \caption{Virtual zone planner (\S~\ref{sec:planner})}
        \label{fig:layout_planning}
    \end{subfigure}
    \vspace{-0.2cm}
    \caption{\textbf{(a)} \textit{Types of collisions: Row-wise collision - Two movements that start on different rows but end on the same row. Diagonal collision - Two movements that start on different columns and end on the same one. Column-wise collision - Two picked-up atoms do not start or end on the same column or row; however, active AOD lasers intersect on an atom that should not be moved.} \textbf{(b)} \textit{Examples of minimal, best, and selected layouts for a circuit with four qubits and at most three concurrent entanglement operations, using single and double row configurations. 
    }}
    \label{fig:movements_planning}
    \Description{}
\end{figure}

The execution time and fidelity of quantum circuits is affected by the physical arrangement of their qubits.
However, understanding how exactly the layout dimensions affect execution remains challenging, because we consider multiple competing objectives --- including fidelity and both spatial and temporal utilization --- the direct function of the layout dimensions is not straightforward.
We define a formula that balances between two boundary layouts: a minimum layout that maximizes QPU utilization for dense packing, and an optimal layout that offers the best performance but occupies more space, thereby reducing co-execution opportunities.
Figure \ref{fig:layout_planning} illustrates examples of these layouts for a four-qubit circuit with a maximum of three concurrent entanglement operations, showing both single and double-row configurations.

\smallskip
\myparagraph{Layout width}
We must select a width that can fit at least the number of qubits $N_q$ in the storage rows $N_r$, making the minimum width $W_{min} = \lfloor N_q/N_r \rfloor \cdot S_s$, where $S_s$ is the storage atom spacing. However, a larger circuit often offers more parallelization opportunities, which benefits performance; for instance, by storing all qubits in a single row, it requires a width $W_s = (N_q-1)\cdot S_s$.
In addition, we may require a greater width $W_e$ to support the largest entanglement operation in the circuit, which Algorithm \ref{alg:layer_spliter} identifies.
The best-performing layout width ($W_{best} = \max(W_s, W_e)$) thus offers maximal parallelism and entanglement.

We select the layout width as a weighted average between the minimum and best-performing with a user-defined performance weight $P_w \in [0,1]$, where $P_w=0$ fully prioritizes QPU utilization (minimal layout) and $P_w=1$ prioritizes circuit performance (best layout). The selected layout width is given by: $W_{selected}=P_w\cdot W_{best}+(1-P_w)\cdot W_{min}$.
After determining $W_{selected}$, the optimized QASM and the selected virtual zone layout for the circuit are transmitted to the target-architecture compiler.

\begin{wrapfigure}[21]{R}{0.5\textwidth}
\begin{algorithm}[H]
\fontsize{8}{9}\selectfont
\caption{Split circuit into layers}
\label{alg:layer_spliter}
\KwData{$circuit$, $window\_size$ \quad \textbf{Result: } $L$}
$L \leftarrow \emptyset$, \(D \leftarrow\) layers(circuit) \Comment{{\color{blue} Convert to DAG layers}} \\
$\text{pred}(n, D)$: Set of predecessors of node $n$ \\
$\text{layer\_compatible}(n, l)$: ($n$ matches gate flag ($S$ or $M$)) $\land$ (all $\text{pred}(n)$ have been executed) \\
\While{$|D| > 0$}{
    $l \leftarrow \emptyset$, $W \leftarrow$ window($D, k, w\_size$) \Comment{{\color{blue} Fetch window of gates to consider for the current layer}}\\
    \ForEach{$layer \in W$ (sorted by gate size)}{
        \ForEach{$n \in layer$}{
            \If{$|l| = 0$ $\lor$ layer\_compatible$(n,l)$}{
                Move $\{n\}$ from $D[k+p]$ to $l$ \Comment{{\color{blue} Remove $n$ from $D$ and add it to the current layer}} \\
                Update $S$ and $M$ depending on gate size 
                continue \\
            }
        }
        $L \leftarrow L\ \cup\ \{l\}$ \Comment{{\color{blue} Add current layer $l$ to the execution layers $L$}} \\
        \lIf{$D[k]$ is empty}{
            \(D \leftarrow D\ \setminus \{D[k]\}\) \Comment{{\color{blue} Remove empty from $D$}}
        }
    }
    \Return{$L$}
}
\end{algorithm}
\end{wrapfigure}
\noindent

\vspace{-12pt}
\subsection{Back-end: Target Compilation}
The final stage of our compiler produces the NA executable, which contains the gate schedule, timing of laser pulses, and zone movements for each execution layer. \projectname remains agnostic to how this is implemented, delegating it to existing back-end compilers, such as ZAC \cite{lin2024reuseawarecompilationzonedquantum} or PacinQo \cite{ludmir2025modeling}, via a compiler abstraction layer.



\smallskip
\myparagraph{Target architecture compiler output}
The compiler produces tuples [$(\text{\qasm}, L)$], one for each circuit compilation. Each tuple contains the result in \qasm with the corresponding virtual zone layout $L$.
Figure \ref{fig:output_grammar}, in Appendix \ref{sec:appendix}, presents the formal definition of the compilation output.
As a case study, we integrate ZAC \cite{lin2024reuseawarecompilationzonedquantum} as a back-end compiler.


\smallskip
\myparagraph{Case Study: ZAIR to \qasm mapping}
Integrating an NA compiler as a target architecture compiler requires mapping its output IR to \qasm. We can directly map most of ZAC's \cite{lin2024reuseawarecompilationzonedquantum} NA instructions to \qasm. \textit{ZAIR} contains four main instructions: (\texttt{init}, \texttt{1qGate}, \texttt{rydberg}, and \texttt{move}), which we map to the corresponding \texttt{init}, \texttt{u3}, \texttt{rydberg}, and \texttt{move} instructions in \qasm by slightly transforming their arguments. We give the full mapping rules in Appendix \ref{sec:zair_to_qasm_formal}.

%% file: tables/qasm_table.tex
\begin{table}[b] 
\centering
\caption{\qasm extensions for NA QPUs.}
\vspace{-6pt}
\fontsize{6.2}{7.2}\selectfont
\begin{tabular}
{p{0.1\textwidth}p{0.13\textwidth}p{0.18\textwidth}p{0.25\textwidth}p{0.20\textwidth}}
\toprule
\textbf{Instruction} & \textbf{Arguments} & \textbf{Description} & \textbf{Pre-condition} & \textbf{Post-condition}  \tabularnewline
\hline
\multirow{1}{*}{\texttt{@init}} & $[(\texttt{x},\texttt{y})_{\texttt{0}...\texttt{n}}]$ \newline $[\texttt{id}_\texttt{0}, ..., \texttt{id}_\texttt{n}]$ & Places and initializes all atoms in the ground state $\ket{0}$ & - & $\forall \mathtt{i}$ $\in$ $[\mathtt{0},\mathtt{n}]$: $\texttt{pos(id}_\texttt{i})$=$(\texttt{x}_\texttt{i},\texttt{y}_\texttt{i})$, \newline $\ket{\Psi_\mathtt{i}} \rightarrow \ket{0}$
\tabularnewline
\midrule
\multirow{1}{*}{\texttt{@move}} & $[(\texttt{x}, \texttt{y})_{\texttt{0}...\texttt{n}}]$ \newline $[(\texttt{x'}, \texttt{y'})_{\texttt{0}...\texttt{n}}]$ & Shuttle logic qubits to input coordinates & Non-overlapping movement contraint (see \ref{section:background}) & $\forall \texttt{i}$ $\in$ $[\texttt{0},\texttt{n}]$: $\texttt{pos(id}_\texttt{i})$=$(\texttt{x}_\texttt{i}, \texttt{y}_\texttt{i})$ \newline $\texttt{state\_is\_preserved}(\texttt{id}_\texttt{i})$
\tabularnewline
\midrule
\multirow{1}{*}{\texttt{@u3}} & $[\texttt{id}_\texttt{0}, ..., \texttt{id}_\texttt{n}]$ \newline [($\theta$, $\phi$, $\lambda)_{\texttt{0}...\texttt{n}}]$ & Apply U3 gates to logic qubits & - & $\forall \texttt{i} \in [\texttt{0},\texttt{n}]$: $\ket{\Psi_\texttt{i}} \rightarrow \texttt{U}_\texttt{3}(\theta,\phi, \lambda)_\texttt{i}\ket{\Psi_\texttt{i}}$ \tabularnewline
\midrule
\multirow{1}{*}{\texttt{@rydberg}} & - & Apply Rydberg pulse to the entanglement zone & - & CZ is applied to all atoms within blockade radius (see \ref{section:background})\tabularnewline
\bottomrule
\end{tabular}
\label{tab:openqasm}
\end{table}

%% file: sections/controller.tex
\section{\projectname (Runtime) Controller}
\label{sec:controller}
The controller serves as the multi-programming back-end, enabling the concurrent execution of multiple circuits produced by our compiler, along with their virtual zone layouts. It receives a list of tiles to be multi-programmed; however, they might not fit in a single execution.
\subsection{Circuit Bundler}
\label{sec:bundler}
The circuit bundler produces execution bins, $B = \{B_1, \dots B_N \}$, where each bin contains a set of circuits $B_j = [c_0, \dots, c_{n_j}]$, that fit within the QPU's space. Naive bundling (\eg{} FIFO) often leads to suboptimal spatial and temporal utilization. Our bundler aims to minimize unused QPU space by selecting tile layouts that maximize hardware utilization, while matching executables with similar depths to avoid idle time caused by different execution durations. The bundler employs a simulated annealing (SA) optimization algorithm to optimize both spatial ($S$) and temporal utilization ($T$). SA minimizes an objective function by iteratively making small modifications in the solution state space. Better modifications are always accepted, while worse ones are accepted based on a "temperature" parameter. This temperature starts high, allowing many suboptimal moves, then gradually decreases (cools down). This approach enables escaping local maxima and finding more optimal global solutions.

\smallskip
\myparagraph{Spatial utilization (${\rho_S}_j$)}
The spatial utilization for an execution bin $B_j$ denotes the used proportion of the QPU area (from 0 to 1).
When $W$ is the total QPU width, and $R$ is the number of storage rows (one or two), the total QPU area is $A_{QPU}=W\cdot R$.
The spatial utilization of bin $B_j$ is \smash{${\rho_S}_j = \sum_{i=0}^{n_j}w_i/{A_{QPU}}$}, where $w_i$ is the width of tile $i$. When ${\rho_S}_j = 1$, the tiles in $B_j$ fit exactly the whole QPU space.

\smallskip
\myparagraph{Temporal utilization (${\rho_T}_j$)}
The temporal utilization captures timing differences between the tiles in bin $B_j$, as a large difference can lead to the QPU being underutilized while deeper executables are running.
Let $d_i$ be the depth of executable $i$, and $D_j = max_{i \in B_j} (d_i)$ be the depth of the deepest executable in bin $B_j$.
The temporal utilization for bin $B_j$ is ${\rho_T}_j = \frac{\sum_{i=0}^{n_j} d_i}{n_j \cdot D_j}$.
When ${\rho_T}_j = 1$, the tiles in $B_j$ all execute for the maximum duration $D_j$.

\smallskip
\myparagraph{Simulated annealing objective} The overall objective function for simulated annealing combines the spatial and temporal utilization for all bins, where we want to maximize $\mathcal{L} = \sum_{j=1}^{N}(\alpha \cdot {\rho_S}_j + (1-\alpha) \cdot {\rho_T}_j) / N$ where $\alpha$ weighs spatial against temporal utilization.
At the extremes, when $\alpha = 1$, we consider only spatial utilization; when $\alpha = 0$, we consider only temporal utilization.

Simulated annealing (SA) starts with an FIFO circuit distribution. In each iteration, we execute one of three actions: \textbf{(1)} Move a tile to a new bin. \textbf{(2)} Swap a tile with one from a different bin. \textbf{(3)} Move a tile to an existing bin. At each iteration, if an action results in a higher utilization $\mathcal{L}$, it is accepted. If it results in lower utilization, the action is accepted depending on the current temperature parameter.

\begin{wrapfigure}[15]{R}{0.44\textwidth}
    \centering
    \includegraphics[width=\linewidth]{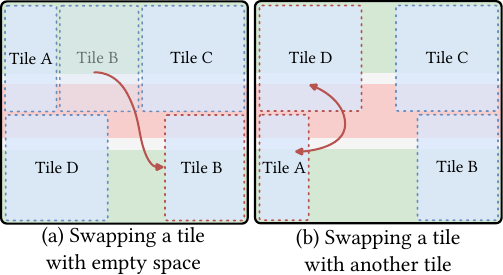}
    \caption{Simulated anneling actions for placement generation (\S~\ref{sec:placement_generator}). \textit{During tile placement, the SA algorithm takes one of two actions: \textbf{(a)} Swapping a tile with empty space or \textbf{(b)} Swapping a tile with another tile, as long as the modified tiles fit in their end positions.}}
    \label{fig:placement}
    \Description{}
\end{wrapfigure}

\smallskip
\myparagraph{AOD lasers constraints}
After bundling all tiles, we must efficiently use shared resources. Particularly, AOD lasers are a critical resource due to their impact on shuttling time. To maximize parallelism and minimize resource contention, movements of atoms must be compatible. This is governed by a set of intra-tile constraints and a set of inter-tile constraints. Inside a tile, AOD movements must be checked for row and column compatibility. On the other hand, across different tiles, AOD movements need to be compatibility-checked on rows, columns, and diagonals in a global coordinate system. Figure \ref{fig:movements} illustrates these rules, which are the base compatibility functions used by the placement generator and orchestrator.

\subsection{Placement Generator}
\label{sec:placement_generator}
After bundling all virtual tiles into several bins, \projectname must place the tiles in each bin $B = \{ t_{1},\ldots,t_{n}\}$ onto the hardware space. This placement is handled by the placement generator and can be modeled as a two-dimensional geometric packing problem with soft objectives, which we refer to as \textit{collision-aware tile placement}.
The QPU is modeled as a grid of size $R \times W$ (i.e.\ $ R$ rows and $W$ columns). The column size is configurable, defining the granularity of the placement search space. The number of rows $R$ is either one or two, depending on whether the QPU is configured with single or double storage. Each tile has a width $w(t_{i}) \in \mathbb{Z}$, measured in grid columns.

AOD lasers impose physical constraints that prevent tiles from executing independently. Following the previously defined movement compatibility rules, each pair of tiles $t_{i}, t_{j}$ has a compatibility cost  $C(t_{i}, t_{j}, p_{i}, p_{j}) \in \mathbb{R}_{\geq 0}$ representing the number of conflicts when placed at positions $p_i$ and $p_j$. $z_{i} \in \{0,1\}$ indicates whether $t_{i}$ is placed, and $x_{i} \in [0, W - w(t_{i})], y_{i} \in [0, R]$ are “anchor position” on the grid. Since an efficient tile placement does not always result in ideal performance, sequentializing some operations may allow more tiles to fit on the QPU. Therefore, we provide parameters $\alpha$ and $\beta$ to control the trade-off between collisions and utilization (a higher $\beta:\alpha$ ratio favors throughput), such that the objective function becomes:
\vspace{-15pt}
\[
c^{\star \star} = min_{p_{1},\ldots,p_{n},z_{1},\ldots,z_{n}} \alpha \cdot \sum_{i < j} z_{i} z_{j} \cdot C(t_{i}, t_{j}; p_i, p_j) - \beta \cdot \sum_{i=1}^{n} z_{i}
\]
\noindent
This problem reduces to VLSI floor planning \cite{7192989}, a well-known NP-hard problem. Therefore, we implement the collision-aware tiling problem heuristically in the placement generator using SA.

\input{tables/algo2}

The placement Algorithm \ref{alg:sa-placer} consists of two parts. First, it finds an initial placement by greedily placing tiles in order of their priority and then by width (placing the smallest tiles first). Then, at each step of the annealing phase, it can opt for one of two actions, as shown in Figure \ref{fig:placement}: tiles can either be swapped with empty space (Figure \ref{fig:placement} (a)) as long as the chosen tile fits the available empty space. Alternatively, a tile can be swapped with another tile (Figure \ref{fig:placement} (b)), as long as both tiles fit the final positions. If this perturbation reduces the cost function, the swap is accepted; otherwise, the probability of it being accepted is proportional to the temperature.

\subsection{Orchestrator}
\label{sec:orchestrator}
Once tiles are placed, the orchestrator schedules NA resources to generate the multi-programmed executable.
Unlike single-circuit compilers, such as ZAC, \projectname must coordinate resources across multiple circuits.
\projectname operates in execution layers $L = \{L_{1},\ldots, L_{k}\}$, each consisting of four phases: \textbf{(1)} movement from the storage to the entanglement zone; \textbf{(2)} Rydberg pulse; \textbf{(3)} movement from the entanglement to the storage zone; and \textbf{(4)} apply single-qubit gates.
To parallelize the forward and reverse movements in each layer, we partition them into sub-rounds of compatible operations.
We model compatibility in a conflict graph $G^{i} = (M^{i}, E)$ where nodes represent NA operations and edges represent conflicts between them.
The lowest number of execution cycles then corresponds to the graph's chromatic number $\chi(G^{i})$.
Since finding $\chi(G^{i})$ is NP-hard, we employ a greedy approach that iteratively removes the maximum independent set until all NA operations are scheduled.

\smallskip
\myparagraph{Single-qubit gates row optimization} When scheduling single-qubit gates, the orchestrator increases parallelization by leveraging a NA hardware capability that allows applying single-qubit gates to targeted atoms in the same row simultaneously (\S~\ref{sec:gate_operations}). However, only $R_Z$ gates can be applied row-wise ($R_Z^R$) on targeted atoms. In contrast, $R_Y$ rotations can only be applied globally ($R_Y^G$) to all the atoms in the array. Given the NA native single-qubit gate set ($R_Z$ and $R_Y$), commonly used $U3(\theta, \phi, \lambda)$ gates, must be decomposed into a $R_Z^R(\phi)R_Y^G(\theta)R_Z^R(\lambda)$ gate sequence. Since the middle $R_Y^G(\theta)$ needs to only affect targeted atoms, it needs to be further synthesized into the $R_Y^G(-\pi/2)R_Z^R(\theta)R_Y^G(\pi/2)$ gate sequence necessary to maintain the state of the non-targeted qubits. The full $U3(\theta, \phi, \lambda)$ gate decomposition would be applied with the following row-optimized gate sequence: $R_Z^R(\phi) + R_Y^G(-\pi/2) + R_Z^R(\theta) + R_Y^G(\pi/2) + R_Z^R(\lambda)$.

%% file: tables/algo2.tex
\begin{wrapfigure}[20]{R}{0.5\textwidth}
\begin{algorithm}[H]
\fontsize{8}{9}\selectfont
\caption{Simulated annealing for collision-aware tile placement}
\label{alg:sa-placer}
\KwData{Tiles $\tau$ with priorities $p_i$, grid size $R \times W$, weights $\alpha, \beta$, cost function $C$}
\KwResult{Tile placement $\{(x_i, y_i, z_i)\}$ minimizing total cost}
Initialize grid $G$ with a greedy placement of tiles sorted by $p_i/w_i$ \\
Initialize temperature $T \gets T_0$, initial placement $P \gets G$ \\
Compute objective $\mathcal{E}(P) \gets \alpha \cdot C(P) - \beta \cdot \sum_i p_i z_i$ \\
\For{$k = 1$ to max iterations}{
    Generate move $P' \gets \text{Perturb}(P)$ \Comment{{\color{blue} Swap or move to empty space}} \\
    \If{$P'$ is feasible (in-bounds, not overlapping)}{
        Compute $\mathcal{E}(P') \gets \alpha \cdot C(P') - \beta \cdot \sum_i p_i z_i$
        $\Delta \gets \mathcal{E}(P') - \mathcal{E}(P)$ \Comment{{\color{blue} Compute new objective to compare}}\\
        \If{$\Delta < 0$ \textbf{or} $\exp(-\Delta / T) > \text{rand()}$}{
            $P \gets P'$ \Comment{{\color{blue} Accept move if new objective is lower}}
        }
    }
    $T \gets \gamma \cdot T$ \Comment{{\color{blue} Reduce temperature}}
}
\Return{Final placement $P$}
\end{algorithm}
\end{wrapfigure}

%% file: sections/consistency-checker.tex
\section{\projectname Checker}
\label{sec:checker}
The checker component ensures that \textit{functional independence} is preserved on the multi-programmed executable. We first formally define that property within the context of neutral-atom multi-programming, and then we explain how the \textit{Checker} verifies functional independence (\Cref{def:independence}).

\subsection{Functional Independence for Multi-Programming}
\label{sec:sequential_consistency}
For each quantum circuit $C_k$ co-executing in an NA multi-programming environment, its functionality is preserved if the containing multi-programmed executable $M$ is equivalent to its original isolated circuit. Recalling from Section \ref{sec:correctness}, the original circuit is defined as: $U_k^{original} = \prod_{i=m_k}^0 U(g_{k_i})$. 

\smallskip
\myparagraph{Multi-programmed executable}
Let $C_k^{original}$ define the isolated original circuit of each $k$ executables in an execution bin. $C_k^{original}$ defines a set of gate-level instructions $G_k = \{G_{k,0},\dots,G_{k,N}\}$. The unitary matrix $U_k^{original}$ defines the overall circuit operation: $\small U_k^{original} = \prod_{i=N}^1 U(G_{k,i})$.

In a multi-programming environment, a $E_k$ executable contains a set of NA instructions applied to the qubits $Q_k = \{q_{k1}, ..., q_{kn}\}$ mapped to the circuit $k$ from a larger set of total qubits $Q$, defined as: $(Q = \bigcup_{k=1}^n Q_{k}) ~\wedge~ (Q_k \cap Q_{k'} = \emptyset: \forall k \neq k')$. Since these NA instructions operate at a lower level than quantum gates, they must first be translated back into equivalent quantum gates, reconstructing a gate-level circuit $C_k^{actual}$. The unitary $U_k^{actual}$ is then derived from the reconstructed gate-level instructions as: $\small U_k^{actual}=U(G'_M)\cdot U(G'_{M-1})\cdot\ ...\ U(G'_2) \cdot U(G'_1)$, where $G'_j$ are the gate-level instructions translated from the NA instructions for executable $E_k$. To ensure functional independence, we must check that the $U_k^{original}$ and $U_k^{actual}$ are equivalent as per \Cref{def:independence}.



\subsection{Functional Independence Checker}
The checker ensures functional independence between the original circuit and the corresponding multi-programmed one. We do this by leveraging quantum circuit reversibility, as described in \Cref{theo:quantum_reversible}. The intuition is that all quantum operations are reversible, allowing us to associate a reverse operation with every forward operation; when combined, they cancel each other out.

\vspace{2pt}
\begin{definition}{Quantum Circuit Reversibility}{theo:quantum_reversible}
All quantum circuits $C$ implement unitary transformations $U$. A transformation is unitary if it satisfies the reversibility property, where $UU^{\dagger} = U^{\dagger}U = I$, where $U^{\dagger}$ is the conjugate transpose (adjoint) of $U$ and $I$ is the identity operator. Therefore, every circuit is also reversible, where $C^{-1}$ implements $U^{\dagger}$.
\end{definition}
\vspace{2pt}

\noindent
To implement this check, we use ZX-diagrams \cite{chen_automata-based_2023}, a representation of quantum circuits based on ZX-calculus. ZX-diagrams encode quantum operations as graphs with colored nodes (Z-spiders and X-spiders) connected by edges representing qubits. The key advantage of ZX-diagrams is their powerful simplification rules, which enable the concatenated diagram to be easily reduced. If the simplified result is an empty graph (representing the identity operation) or a global phase, the circuits are functionally equivalent, as per \Cref{def:independence}, confirming their functional independence.
By concatenating the ZX-diagram of the original circuit with the corresponding inverted multi-programmed version and applying a set of ZX-diagram simplification passes, it allows the checker to infer functional equivalence between both circuits, as defined in Section \ref{sec:sequential_consistency}.
Figure \ref{fig:checker} gives an example of the Checker workflow, of which we explain the steps in more detail below:

\begin{figure}
    \centering
    \includegraphics[width=\textwidth]{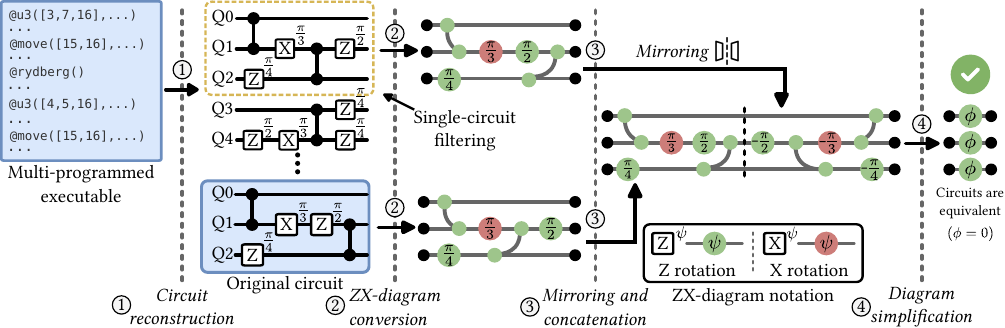}
    \caption{Checker workflow (\S~\ref{sec:checker}). \textit{The \textit{Checker} takes as input the original circuit from the Compiler and the multi-programmed executable from the Controller. \textbf{(1)} The executable is reconstructed into a quantum circuit and constrained to qubit $Q_k$ corresponding to the original circuit. \textbf{(2)} Both circuits are converted into their respective ZX-diagrams. \textbf{(3)} The ZX-diagram of the original circuit is concatenated with the mirroring of the ZX-diagram of the reconstructed circuit, swapping the signs of the rotation angles. \textbf{(4)} Finally, the concatenated ZX-diagram is simplified, which results in an empty circuit, meaning that both circuits are equivalent.}}
    \Description{}
    \label{fig:checker}
\end{figure}

\smallskip
\myparagraph{\#1: Circuit reconstruction}
NA QPUs employ a relatively simple instruction set at the algorithmic level, which facilitates the recovery of the circuit semantics for each program in the output executable. The checker statically analyzes which gates are executed by maintaining a virtual state of each atom and checking their positions for gate execution (e.g.\ within the Rydberg zone). In Section \ref{sec:sequential_consistency}, this translation process was abstracted away in the notation $U(G)$. In Figure \ref{fig:checker}, the example starts with a set of \qasm instructions that can be reconstructed into a circuit on the right. From this circuit, qubits $Q_k$ are represented by the first three qubits inside the yellow box.

\smallskip
\myparagraph{\#2: ZX-diagram conversion}
An interpretation function $\llbracket \cdot \rrbracket: \textbf{Circuit} \rightarrow \textbf{Set}(\textbf{ZX})$ translates both the original $C_k^{original}$ and the multi-programmed circuit $C_k^{actual}$ into ZX-diagrams using the standard circuit to ZX translation (provided by \textit{PyZX} \cite{noauthor_pyzx_nodate}). We aim to show that these diagrams are equivalent and represent the same unitary transformations. In Figure \ref{fig:checker}, the circuits are converted to their ZX-diagrams in step 2. In ZX-diagram notation, $X$ rotations are represented with a red node, while $Z$ rotations are represented with a green node, both with the respective rotation angles. $CZ$ gates are symmetrical and represented with two green nodes on both interacting qubits.

\smallskip
\myparagraph{\#3: Mirroring and concatenation}
We create the ZX-diagram representing the adjoint of $C_k^{actual}$ with a ``mirroring'' operation, where the inputs become outputs (and vice-versa) and negating the phases $(\alpha \mapsto -\alpha)$ for all Z- and X-spiders. The two diagrams $C_k^{original}$ and $(C^{actual}_k)^{\dagger}$ are then concatenated together. This is represented on step 3 of Figure \ref{fig:checker}.

\smallskip
\myparagraph{\#4: Diagram simplification}
Finally, a ZX simplification pass is run on the resulting diagram: if the resulting diagram is either the identity operation or $Z$ rotation on all qubits with the same phase representing a global phase, then we can deduce \smash{$U_k^{original} \cdot (U^{actual}_k)^{\dagger} = e^{i \phi} I$}; the compiled circuit is functionally equivalent to the input. This process is always possible and will terminate as the ZX-calculus is sound and complete over the gate-set fragment $\{X, Y, \text{CZ}\}$. In step 4 of Figure \ref{fig:checker}, the concatenated diagram is reduced to the identity diagram (where global phase $\phi=0$).

%% file: sections/evaluation.tex
\section{Evaluation}
\label{sec:evaluation}
We structure the evaluation in three core parts: \textbf{full system} end-to-end analysis (\S~\ref{sec:end_to_end_eval}), \textbf{compiler} analysis (\S~\ref{sec:compiler_eval}), and (runtime)  \textbf{controller} analysis (\S~\ref{sec:controller_eval}).

\subsection{Experimental Methodology}
\myparagraph{Baselines} Across all evaluations, we compare \projectname against ZAC \cite{lin2024reuseawarecompilationzonedquantum} and PachinQo \cite{ludmir_modeling_2024}, the state-of-the-art compilers for zoned NA architectures.

\smallskip
\myparagraph{Benchmarks} We use 11 benchmarks from two standard benchmark suites~\cite{li2023qasmbench, quetschlich_mqt_2023} (see Table \ref{tab:benchmarks} in the Appendix \ref{sec:appendix}). For fairness, we use benchmarks similar to those used by the baselines.

\smallskip
\myparagraph{Fidelity model}
\label{sec:fid_model}
We employ a widely used model to estimate fidelity \cite{lin2024reuseawarecompilationzonedquantum}. The fidelity model considers four main sources of error: one-qubit gate error ($E_1$), two-qubit gate error ($E_2$), atom transfer error ($E_{trans}$), and decoherence time ($T_2$). We compute the resulting fidelity $f$ as:
$$
f = (E_1)^{n_1} \cdot (E_2)^{n_2} \cdot (E_{trans})^{n_{trans}} \cdot \prod_{q \in Q} \exp\left(-\frac{t_q}{T_2}\right)
$$
\vspace{-7pt}

\input{tables/fidely_execution_means}
\myparagraph{QPU hardware setup} We evaluate the QPU architecture with single and double storage zones. The QPU hardware setup and parameters are detailed in Table~\ref{tab:hardware_parameters} in the Appendix \ref{sec:appendix}. We conservatively estimate the initialization overhead to be 82 ms for a 280-qubit QPU \cite{labuhn_realizing_2016, wintersperger_neutral_2023}.

\begin{figure}[t]
    \centering
    \includegraphics[width=\linewidth]{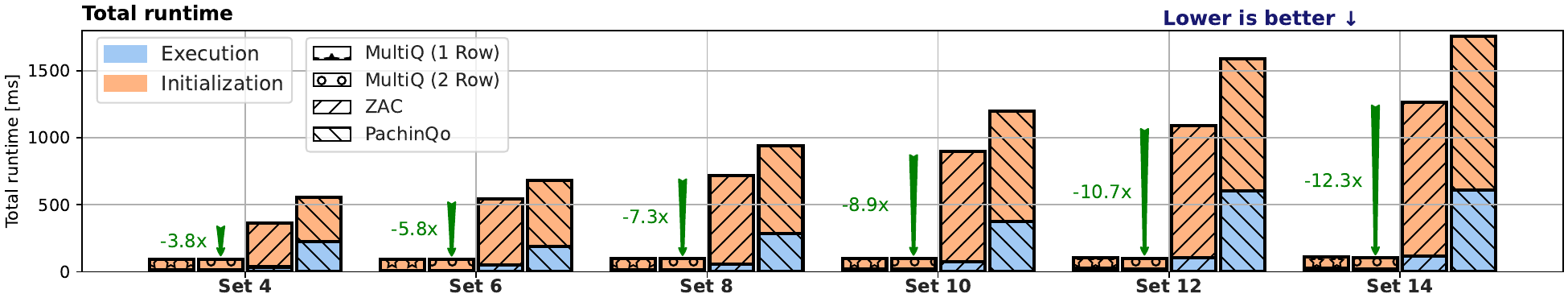}
    \vspace{-16pt}
    \caption{\textbf{RQ\#1:} End-to-end total runtime evaluation comparing MultiQ, ZAC, and PachinQo.}
    \Description{}
    \label{fig:e2e_durations}
    \vspace{-4pt}
\end{figure}

\myparagraph{Metrics} We evaluate \projectname across five metrics: \textbf{(1)} Fidelity (\S~\ref{sec:fid_model}); \textbf{(2)} circuit execution time; \textbf{(3)} total duration, comprising the circuit execution time and the QPU initialization time; \textbf{(4)} spatial utilization; and finally, \textbf{(5)} temporal utilization (\S~\ref{sec:bundler}).

\subsection{Full System Evaluation}
\label{sec:end_to_end_eval}
{\bf RQ1:} \textit{What is \projectname's multi-programming runtime improvement?} This assesses the overall effectiveness of the system at minimizing the total runtime, thus increasing the QPU throughput. We compare \projectname's multi-programmed sets of 4 to 14 circuits to ZAC's sequential (solo) execution. We randomly bundled the benchmarks, avoiding duplicate circuits within the same set.

\myparagraph{Total runtime improvement}
Figure \ref{fig:e2e_durations} shows \projectname's ability to reduce total system runtime. By multi-programming 14 circuits, \projectname achieves a runtime reduction of up to $12.3\times$. As discussed in the preliminary evaluation (Section \ref{sec:introduction}), initialization time represents a significant overhead of the total runtime. When executing 14 circuits, \projectname reduces the initialization overhead from 14 QPU initializations to just one, which represents a significant improvement that seems to account for the main runtime gains in Figure \ref{fig:e2e_durations}. Additionally, multi-programming enables parallel execution of these 14 circuits with minimal runtime increase, as concurrent instruction contention remains negligible compared to the overhead of sequential execution.

\begin{figure*}[t]
    \centering
    \includegraphics[width=\textwidth]{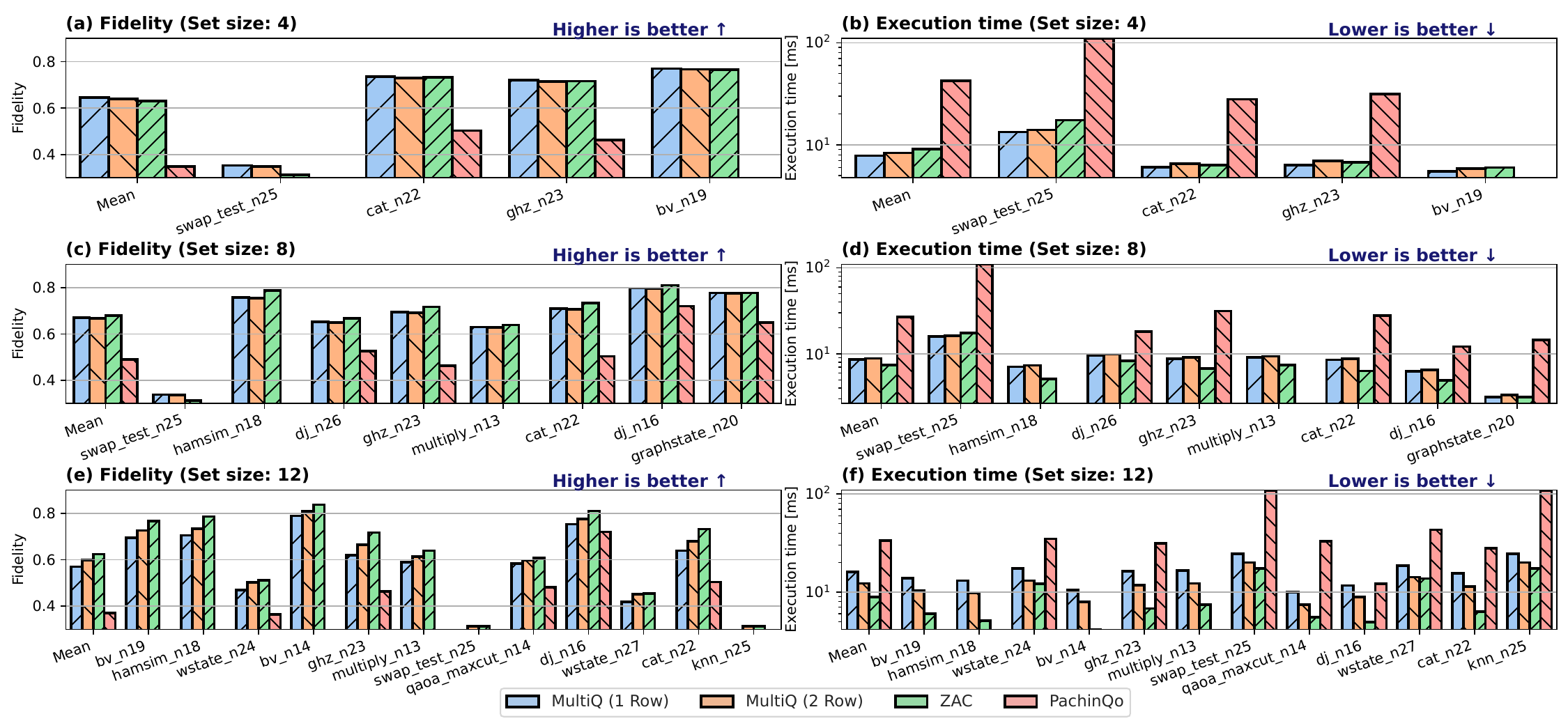}
    \vspace{-15pt}
    \caption{\textbf{RQ\#2:} End-to-end evaluation: fidelity and circuit duration (\S~\ref{sec:end_to_end_eval}). \textit{\textbf{(a)} Fidelity of each benchmark co-executed by \projectname vs ZAC and PachinQo solo executions. \textbf{(b)} Circuit execution time of each benchmark as executed by \projectname, ZAC and PachinQo.}}
    \Description{}
    \label{fig:e2e_plot_detailed}
    \vspace{-10pt}
\end{figure*}

\smallskip
{\bf RQ2:} \textit{What is \projectname's multi-programmer's fidelity performance with respect to solo execution?} We next assess \projectname's effectiveness w.r.t fidelity and circuit execution time, comparing co-executing circuits vs solo circuit execution. We use the same methodology as the previous research question.

\myparagraph{Fidelity performance}
Figure \ref{fig:e2e_plot_detailed} shows the fidelity and execution time for three evaluated sets: 4, 8, and 12 circuits. On the left side, it shows the fidelity values over the benchmarks included in the sets, while on the right side, it shows the execution time. We can see that for larger sets, \projectname shows a slight fidelity drop, averaging $-2.6$\% for a set of 12 circuits, when compared to isolated ZAC's execution. This drop is accompanied by an increase in execution time, averaging $+3.4$ ms ($+38\%$) on a set of 12 circuits, which suggests a slight increase in decoherence error resulting from instruction contention between co-executed circuits. Table \ref{tab:means} presents a broader view, showing multi-programmed sets with 4 to 14 circuits, where the same behaviour occurs, especially for larger circuit sets. Overall, \projectname largely preserves single-circuit fidelity, from a 1.33\% increase in fidelity with a set of 4 circuits, to a small 3.51\% fidelity drop when co-executing 14 circuits simultaneously. Pachinqo \cite{ludmir_modeling_2024} achieves lower fidelity results compared to both ZAC and \projectname.

\subsection{Compiler Evaluation}
\label{sec:compiler_eval}
{\bf RQ3:} \textit{What is the effect of the virtual zone planner on trading off circuit performance and QPU utilization?} This evaluation explores the virtual layout planner at trading off the circuit's performance and QPU utilization. We evaluate this trade-off by running single circuits, varying the planner's performance weight, and measuring decoherence error and free QPU space, which helps to visualize how much QPU space remains available. We select six benchmarks with distinct tradeoff behaviors.

\begin{figure*}[t]
    \centering
    \includegraphics[width=\textwidth]{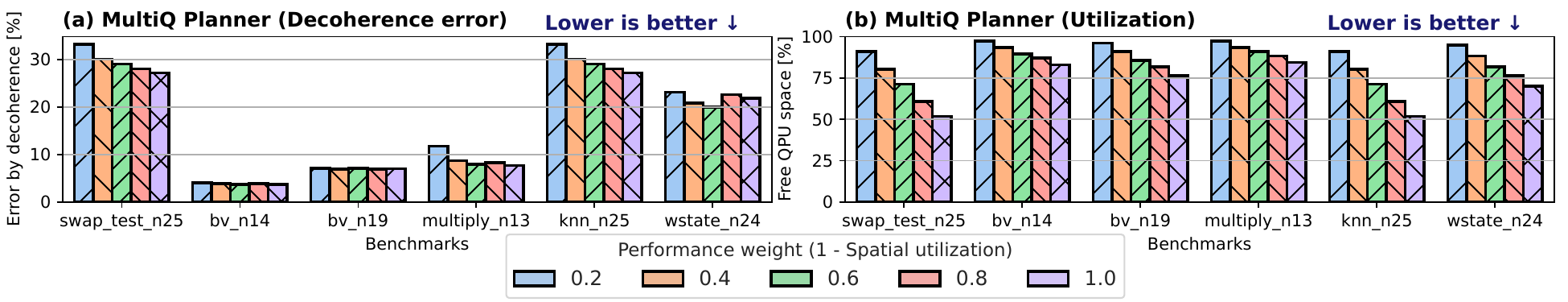}
    \vspace{-14pt}
    \caption{\textbf{RQ\#3:} Virtual layout planner (\S~\ref{sec:compiler}) evaluation. {\em \textbf{(a)} Shows the effect of decoherence error on different benchmarks by increasing the performance weight on the formula in \ref{sec:planner}. \textbf{(b)} Shows the effect on free QPU utilization.}}
    \Description{}
    \label{fig:planner_eval}
    \vspace{-1mm}
\end{figure*}

\myparagraph{Analysis of the virtual layout planner results}
Figures \ref{fig:planner_eval} (a) and (b) show the tradeoff of increasing the performance weight on the layout planning formula explained in Section \ref{sec:planner}, as higher values in performance weight lead the narrower layouts.
Figure \ref{fig:planner_eval} (a) shows just a slight decrease in decoherence error, in most benchmarks, at most a $5$\% decrease in performance weights of 0.2 and 1.0. On the other hand, Figure \ref{fig:planner_eval} (b) shows a sharp loss of free QPU space, as wider layouts reduce the number of circuits that can fit in a single execution bundle, from an average of 92\% QPU spatial utilization at 0.8 spatial utilization weight to an average of 65\% QPU utilization at 0.2 utilization weight. The decrease in QPU utilization is especially accentuated for larger benchmarks. In conclusion, setting a spatial utilization weight on the higher end, between 0.6 and 0.8, produces narrower layouts (see Figure \ref{fig:layout_planning}), allowing \projectname to increase quantum circuit throughput with minimal sacrifice in error due to decoherence, for example, setting a spatial utilization weight at 0.6 (0.4 performance weight) sacrifices only a average of 2\% decoherence fidelity (comparing with a 1.0 performance weight) but achieves an averages of 20\% higher QPU spatial utilization.


\subsection{Controller Evaluation}
\label{sec:controller_eval}
\textbf{RQ4:} \textit{What is the effect of the circuit bundler on maximizing spatial and temporal QPU utilization?} Here, we investigate the effectiveness of the circuit bundler at grouping quantum circuits into execution bins. We compare the circuit bundler's effectiveness against a FIFO approach using sets of 6 to 14 circuits. We set up the planner to produce larger zone layouts (performance-focused layout planning), where one execution cycle would not fit all the circuits.

\begin{figure*}[t]
    \centering
    \includegraphics[width=\textwidth]{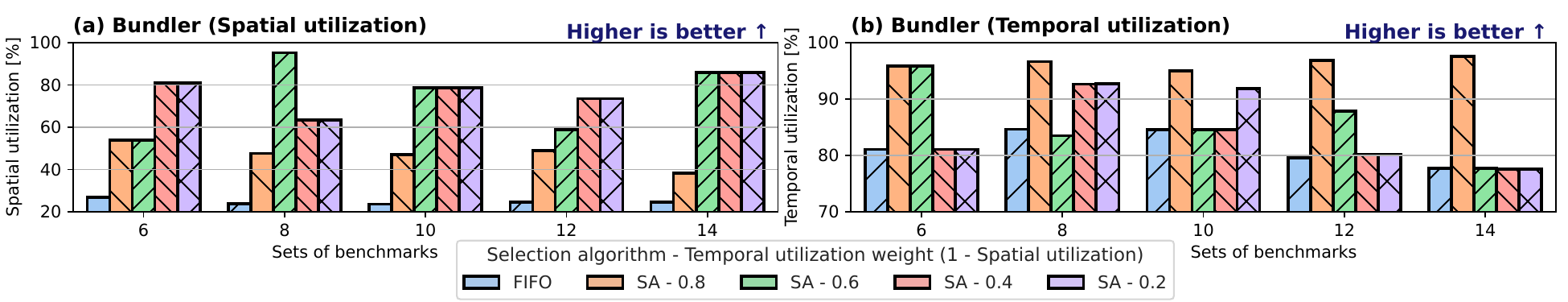}
    \vspace{-12pt}
    \caption{\textbf{RQ\#4:} Circuit bundler (\S~\ref{sec:compiler}) evaluation. {\em \textbf{(a)} Shows the circuit bundler results of QPU spatial utilization for a FIFO and simulated annealing (with different cost weights on temporal utilization) algorithms. \textbf{(b)} Shows the effects of the different bundling of the algorithms on temporal utilization.}}
    \Description{}
    \label{fig:bundler_eval}
    \vspace{-2mm}
\end{figure*}

\myparagraph{Analysis of the circuit scheduler results}
Figure \ref{fig:bundler_eval} (c) shows a sharp increase in QPU spatial utilization compared to a simple FIFO bundling algorithm, particularly at higher spatial weight values, up to 80\% increase on a set of 14 circuits with 0.8 spatial utilization weight (0.2 temporal utilization weight). This is expected: as the algorithm prioritizes spatial optimization, it bundles circuits with better hardware-fitting layouts. On the other hand, in Figure \ref{fig:bundler_eval} (d), average temporal utilization decreases with lower temporal utilization weights, averaging a 10\% decrease between 0.8 and 0.2 temporal utilization weights, which shows a lower trade-off behavior than expected. This suggests that maximizing spatial efficiency does not necessarily compromise temporal utilization of bundled circuits. We hypothesize that this relationship depends vastly on the depth of the pool of input circuits; for example, when all circuits have similar runtimes, adjusting the spatial-to-temporal weight ratio has minimal impact on temporal metrics but a large impact on space utilization. Overall, the simulated annealing bundling algorithm achieves $3\times$ higher spatial QPU utilization and improved temporal utilization compared to the FIFO approach.

\smallskip
{\bf RQ5:} \textit{What is the effect of the controller at layout placement and independent circuit execution parallization?} We compare the controller's efficiency in layout placement and execution parallelization with a naive circuit-merging approach. We select random sets of benchmarks with an increasing number of circuits (from 4 circuits to 14 circuits). Conversely to RQ1 and RQ2, instead of running the circuits in the set sequentially, the baselines run a single quantum circuit that is the result of merging all the quantum circuits in the set in parallel.

\begin{figure*}[t]
    \centering
    \includegraphics[width=\textwidth]{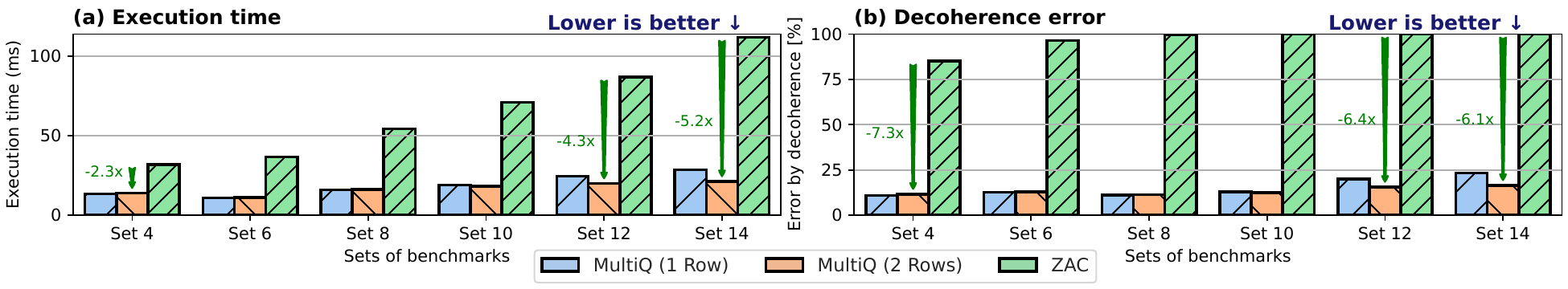}
    \vspace{-16pt}
    \caption{\textbf{RQ\#5:} Controller's evaluation of execution time and decoherence error on increasing number of circuits (\S~\ref{sec:controller_eval}). \textit{\textbf{(a)} Execution time for \projectname's controller and ZAC compiler. \textbf{(b)} Decoherence error results.}}
    \Description{}
    \label{fig:controller_eval}
    \vspace{-1mm}
\end{figure*}

\myparagraph{Parallelization performance}
Figure \ref{fig:controller_eval} (a) shows a sharp reduction in the total circuit execution time by \projectname's controller, compared to ZAC, up to $5.2\times$. The reduction in execution time results in a strong decrease in decoherence error from almost ZAC's 100\% on sets of 6 circuits or more to approximately 15\% by \projectname (Figure \ref{fig:controller_eval} (b)). This is due to the fact that ZAC's compilation approach is limited by space constraints and is unable to properly place independent circuits, which leads to high shared resource contention, longer runtimes, and thus higher fidelity loss due to decoherence errors. Overall, \projectname's placement and parallelization approaches achieve large improvements in total execution time and decoherence errors compared to a naive circuit merging solution.

%% file: tables/fidely_execution_means.tex
\begin{wrapfigure}[12]{r}{0.6\textwidth}
    \vspace{-9pt}
    \caption{\textbf{RQ \#1:} Fidelity and circuit execution time means for different multi-programming sets.}
    \fontsize{7}{8}\selectfont
    \begin{tabular}{p{0.03\linewidth}|p{0.22\linewidth}|p{0.065\linewidth}|p{0.065\linewidth}|p{0.065\linewidth}|p{0.065\linewidth}|p{0.065\linewidth}|p{0.065\linewidth}|}
    \toprule
    & \textbf{Compiler} & \textbf{Set 4} & \textbf{Set 6} & \textbf{Set 8} & \textbf{Set~10} & \textbf{Set~12} & \textbf{Set~14}\\
    \midrule
    \multirow{4}{*}{\rotatebox[origin=c]{90}{\textbf{Fidelity (\%)}}} & \textbf{ZAC} & \cellcolor{red!25} 63.18 & 65.02 & 67.99 & 67.81 & 62.37 & 63.77 \cellcolor{green!25}\\ \cmidrule(lr){2-8}
                              & \textbf{MultiQ~(1~Row)} & \cellcolor{green!25} 64.51 & 64.46 & 66.94 & 65.77 & 57.00 & 55.07 \\ \cmidrule(lr){2-8}
                              & \textbf{MultiQ~(2~Row)} & \cellcolor{green!25} 64.05 & 64.35 & 66.71 & 66.12 & 59.81 & 60.26 \cellcolor{red!25} \\
                              \cmidrule(lr){2-8}
                              & \textbf{PachinQo} & 34.93 & 40.69 & 49.07 & 45.93 & 37.10 & \cellcolor{red!50} \cellcolor{red!50} 41.45\\
    \midrule
    \multirow{4}{*}{\rotatebox[origin=c]{90}{\textbf{Time (ms)}}} & \textbf{ZAC} & \cellcolor{red!25} 9.13 & 8.66 & 7.43 & 7.65 & 8.91 & \cellcolor{green!25} 8.45 \\ \cmidrule(lr){2-8}
                              & \textbf{MultiQ~(1~Row)} & \cellcolor{green!25} 7.82 & 8.90 & 8.57 & 10.04 & 16.07 & 19.01 \\ \cmidrule(lr){2-8}
                              & \textbf{MultiQ(2~Row)} & \cellcolor{green!25} 8.32 & 9.01 & 8.82 & 9.62 & 12.27 & \cellcolor{red!25} 12.83 \\
                              \cmidrule(lr){2-8}
                              & \textbf{PachinQo} & \cellcolor{red!50} 56.33 & 31.94 & 35.64 & 37.60 & 50.21 & \cellcolor{red!50} 43.43 \\
    \hline
    \bottomrule
    \end{tabular}
    \label{tab:means}
\end{wrapfigure}

%% file: sections/related_work.tex
\section{Related Work}
\label{section:related_work}

\myparagraph{Quantum compilers}
Quantum compilers translate high-level quantum circuits into operations that can be executed by quantum hardware, and their development is an active area of research. There exist numerous compilers for superconducting qubits \cite{qiskit-transpiler, hua2023caqr, gushu2019tackling, prakash2019noise, swamit2019not, chi2021time, yunong2019optimized, tannu2019ensemble, liu2022not, murali2020software}, trapped ions \cite{gorenland2020signal, saki2022muzzle, maslov2017basic, kreppel2023quantum, schmale2022backend, chang2025quantum}, and photonic quantum computers \cite{zilk2022a, zhang2023oneq, zhangoneperc2024}. However, they are designed around specific features/challenges of those architectures and are not suitable for NAs.

\myparagraph{NA compilers}
Existing compilers for NA architectures either target static hardware or support limited dynamic capabilities such as qubit shuttling or zoned layouts \cite{baker2021exploiting, patel2022geyser, patel2023graphine, wang_atomique_2024, tan2022qubit, tan2025compilation, ludmir_parallax_2024, tan2024compilingquantum, kirmenis2025weaver, lin2024reuseawarecompilationzonedquantum, stade2024an, jang2025qubit, ludmir2025modeling}. However, none fully exploit the range of NA features for performance, or support both zoning and multi-programming. In contrast, \projectname leverages all state-of-the-art NA capabilities, including zoned architectures and multi-programming.

\myparagraph{NA controllers} A NA controller translates the quantum compiler’s output into the precise control signals needed to manipulate individual neutral atoms~\cite{anand_dual-species_2024, stein_multi-target_2025, zhang_high_2024, steinert_spatially_2023}. Related research focuses on accelerating atom rearrangements with new algorithms \cite{wang_accelerating_2023} or hardware, such as FPGAs \cite{guo_fpga-accelerated_2024}. Our project, \projectname, builds upon this with support for custom atom layouts.

\myparagraph{Multi-programming QPUs}
Although multi-programming has been explored for superconducting qubits \cite{das2019a, liu2021qucloud, giortamis2025qos}, multi-programming in NAs faces unique challenges (\S~\ref{sec:motivation}), rendering the aforementioned works non-applicable. Unfortunately, no multi-programming work exists on NA. 

\myparagraph{Quantum HW-SW co-design}
Quantum HW-SW co-design has been explored across architecture design, error correction, and distributed quantum computing to improve application fidelity and optimize quantum resources \cite{stein2025hetec, lin2024codesign, wang2024optimizing, ang2024arquin, li2021on, stein2023hetarch}. Most of these efforts, excluding PachinQo \cite{ludmir2025modeling}, focus on superconducting QPUs and single-program scenarios. In contrast, \projectname advances HW-SW co-design by addressing the challenges of multi-programming for NA QPUs.

\myparagraph{Formal methods in quantum computing} Formal methods in quantum computing cover formal verification (ensuring circuits work as intended \cite{leite_ramalho_testing_2025, abdulla_verifying_2024, noauthor_efficient_nodate}) and equivalence checking (confirming two circuits are functionally identical). Equivalence checking is QMA-hard \cite{janzing_non-identity-check_2005} and computationally expensive, and implementations exist on accelerators such as GPUs \cite{osama_parallel_2025}. \projectname addresses this using ZX-calculus \cite{peham_equivalence_2022, duncan_graph-theoretic_2020} and is the first to bring this capability to a multiprogramming environment.

%% file: sections/conclusion.tex
 \section{Conclusion}
\label{section:conclusion}

We present \projectname, a compiler-controller co-design that enables high-throughput, fidelity-aware multi-programming on NA QPUs. \projectname partitions and maps multiple circuits to non-overlapping QPU regions, co-optimizing for utilization, fidelity, and latency, while ensuring correctness by checking functional independence. Implemented on top of Qiskit and ZAC, our evaluation shows that \projectname improves QPU throughput by 5.4$\times$ to 21$\times$ with minimal fidelity loss (0–2.7\%) when running up to 10 circuits concurrently.

\myparagraph{Artifact} \projectname will be publicly available as an  open-source project. 
 
\myparagraph{Appendix} The appendix contains \qasm grammar, mapping rules from the ZAC IR to the \qasm IR, the hardware experimental setup,  and the benchmark details.

%% file: sections/appendix.tex
\newpage
\appendix
\section{Appendix}
\label{sec:appendix}

\subsection{\qasm Grammar}

Figure \ref{fig:grammar}, formalizes the \qasm grammar in EBNF format. Each program begins with an optional version declaration, followed by a body composed of one or more statements or scopes. Each statement can start with a \textit{pragma} or an \textit{annotation}, which adds information to the subsequent statement body. The statement body then contains common QASM quantum operations (\eg{} qubit initialization, operations on qubits, etc.). The supported annotations, introduced in \qasm and detailed in Table \ref{tab:openqasm}, are inserted before the relevant statement bodies and extend the base QASM language with neutral-atom-specific functionalities.
\vspace{-5pt}

\input{tables/grammar}

\begin{figure}
    \centering
    \begin{subfigure}[c]{0.53\textwidth}
        \begin{tcolorbox}[colback=white, colframe=black, boxrule=0.3pt,
        arc=2pt, left=4pt, right=4pt, top=4pt, bottom=4pt]
        \fontsize{8}{9}\selectfont
        \setlength{\grammarparsep}{6pt plus 1pt minus 1pt}
        \vspace{0.3cm}
        \begin{grammar}
        <\textbf{\textcolor{\grammarRuleDefColor}{output}}> ::= \{ <tileInfo> [ \textcolor{\grammarTerminalColor}{‘","’} <tileInfo>] \}
    
        <\textbf{\textcolor{\grammarRuleDefColor}{tileInfo}}> ::= <\qasm>\textcolor{\grammarTerminalColor}{‘","’} <virtualZoneLayout>
    
        <\textbf{\textcolor{\grammarRuleDefColor}{virtualZoneLayout}}> ::= <qpuVariables>\textcolor{\grammarTerminalColor}{‘","’} <storageVariables> \textcolor{\grammarTerminalColor}{‘","’} <entanglementVariables>
    
        <\textbf{\textcolor{\grammarRuleDefColor}{storageVariables}}> ::= \{<storageVariable> [\textcolor{\grammarTerminalColor}{‘","’}<storageVariable>]\}
    
        <\textbf{\textcolor{\grammarRuleDefColor}{entanglementVariables}}> ::= \{<entanglementVariable> [\textcolor{\grammarTerminalColor}{‘","’}<entanglementVariable>]\}
    
        <\textbf{\textcolor{\grammarRuleDefColor}{qpuVariables}}> ::= <width>\textcolor{\grammarTerminalColor}{‘","’}<height>\textcolor{\grammarTerminalColor}{‘","’}<nqubits>\textcolor{\grammarTerminalColor}{‘","’} <nAODs>\textcolor{\grammarTerminalColor}{‘","’}<zoneSeparation>

        <\textbf{\textcolor{\grammarRuleDefColor}{storageVariable}}> ::= <width>\textcolor{\grammarTerminalColor}{‘","’}<height>\textcolor{\grammarTerminalColor}{‘","’}<position>

        <\textbf{\textcolor{\grammarRuleDefColor}{entanglementVariable}}> ::= <width>\textcolor{\grammarTerminalColor}{‘","’}<height>\textcolor{\grammarTerminalColor}{‘","’} <position>

        <\textbf{\textcolor{\grammarRuleDefColor}{position}}> ::= \textcolor{\grammarTerminalColor}{‘"("’} \syntleft\textcolor{\grammarTypeColor}{float}\syntright \textcolor{\grammarTerminalColor}{‘","’} \syntleft\textcolor{\grammarTypeColor}{float}\syntright \textcolor{\grammarTerminalColor}{‘")"’}

        <\textbf{\textcolor{\grammarRuleDefColor}{zoneSeparation}}> ::= \textcolor{\grammarTerminalColor}{‘"("’} \syntleft\textcolor{\grammarTypeColor}{float}\syntright \textcolor{\grammarTerminalColor}{‘","’} \syntleft\textcolor{\grammarTypeColor}{float}\syntright \textcolor{\grammarTerminalColor}{‘")"’}
        \end{grammar}
        \end{tcolorbox}
        \caption{\textit{Target architecture compilation} output}
        \label{fig:output_grammar}
        \vspace{4pt}
    \end{subfigure}
    \begin{subfigure}[c]{0.46\textwidth}
        \begin{tcolorbox}[colback=white, colframe=black, boxrule=0.3pt, arc=2pt, left=4pt, right=4pt, top=4pt, bottom=4pt]
        \fontsize{8}{9}\selectfont
        \vspace{-0.2cm}
        \begin{align*}
        <\grammarOp{init}> ::=  &\{\grammarArg{init\_locs}: \grammarType{list[(x,y)]}\} \\
        <\grammarOp{1qGate}> ::= &\{\grammarArg{unitary}: \grammarType{u3}, \\
        &\grammarArg{init\_locs}: \grammarType{list[(x,y)]}\} \\
        <\grammarOp{rydberg}> ::= &\{\grammarArg{zone\_id}: \grammarType{int}\} \\
        <\grammarOp{move}> ::= &\{\grammarArg{row\_id}: \grammarType{list[int]}, \\
        &\grammarArg{row\_y\_begin}: \grammarType{list[float]}, \\
        &\grammarArg{row\_y\_end}: \grammarType{list[float]}, \\
        &\grammarArg{col\_id}: \grammarType{list[int]}, \\
        &\grammarArg{col\_x\_begin}: \grammarType{list[float]}, \\
        &\grammarArg{col\_x\_end}: \grammarType{list[float]}\}
        \end{align*}
        \end{tcolorbox}
        \caption{\textit{ZAIR} \cite{lin2024reuseawarecompilationzonedquantum} IR}
        \label{fig:zair_grammar}
    \end{subfigure}
    \vspace{-0.3cm}
    \caption{\textbf{(a)} Formal grammar of the output of the \textit{Target architecture compilation} in EBNF format. The formal definition of the \qasm grammar is defined in \ref{sec:front_end}. \textbf{(b)} Instruction definitions of ZAC's IR, named ZAIR \cite{lin2024reuseawarecompilationzonedquantum}}
    \label{fig:movements_planning}
    \Description{}
\end{figure}

\subsection{ZAIR to \qasm mapping}
\label{sec:zair_to_qasm_formal}
The mapping from ZAIR \cite{lin2024reuseawarecompilationzonedquantum} instructions to \qasm can be formally defined through the following functions. For reference, Figure \ref{fig:zair_grammar} shows the ZAIR \cite{lin2024reuseawarecompilationzonedquantum} instruction list.
\[
T_{\text{init}}: \text{init}_{\text{ZAIR}}(\mathit{init\_locs}) \mapsto \text{init}_{\text{\qasm}}(\mathit{init\_locs})
\]
\[
T_{\text{1qGate}}:\ \text{1qGate}_{\text{ZAIR}}(u3, init\_locs) \mapsto \text{u3}_{\text{\qasm}}(init\_locs, [\forall i \in u3 (u3.x, u3.y, u3.z)])
\]
\[
T_{\text{rydberg}}: \text{rydberg}_{\text{ZAIR}}(\mathit{zone\_id}) \mapsto \text{rydberg}_{\text{\qasm}}(\mathit{zone\_id})
\]
\[
T_{\text{move}}: \text{move}_{\text{ZAIR}}(zone\_id,row\_id,row\_y\_begin,row\_y\_end,col\_id,col\_x\_begin,col\_x\_end)\] \[\mapsto \text{move}_{\text{PhysIR}}(row\_y\_begin,col\_x\_begin,row\_y\_end,col\_x\_end)
\]

\subsection{Benchmarks and Experimental setup}
The list of used benchmarks and experimental setup is listed in Tables \ref{tab:benchmarks} and \ref{tab:experimental_setup}, respectively.

\input{tables/experimental_setup}

%% file: tables/grammar.tex
\begin{figure}[th]
\begin{tcolorbox}[colback=white, colframe=black, boxrule=0.3pt,
  arc=2pt, left=4pt, right=4pt, top=4pt, bottom=4pt]
    \fontsize{8}{9}\selectfont
    \setlength{\grammarparsep}{6pt plus 1pt minus 1pt}
    \vspace{0.3cm}
    \begin{grammar}
    <\textbf{\textcolor{\grammarRuleDefColor}{program}}> ::= <version> \textcolor{\grammarSymbolColor}{?} <statementOrScope>\textcolor{\grammarSymbolColor}{*}
    
    <\textbf{\textcolor{\grammarRuleDefColor}{version}}> ::= \textcolor{\grammarTerminalColor}{‘"OpenQASM"’} <versionSpecifier> \textcolor{\grammarTerminalColor}{‘";"’}
    
    <\textbf{\textcolor{\grammarRuleDefColor}{statementOrScope}}> ::= <statement> \, \textcolor{\grammarSymbolColor}{|} \, <scope>
    
    <\textbf{\textcolor{\grammarRuleDefColor}{scope}}> ::= \textcolor{\grammarTerminalColor}{‘"\{"’} <statementOrScope> \textcolor{\grammarSymbolColor}{*} \textcolor{\grammarTerminalColor}{‘"\}"’}
    
    <\textbf{\textcolor{\grammarRuleDefColor}{statement}}> ::= <pragma>
        \textcolor{\grammarSymbolColor}{\alt} <annotation>\textcolor{\grammarSymbolColor}{*} \textcolor{\grammarSymbolColor}{(}
            \\ \textcolor{\grammarSymbolColor}{\textbar} \quad <ioDeclarationStatement>
            \\ \textcolor{\grammarSymbolColor}{\textbar} \quad <gateStatement>
            \\ \textcolor{\grammarSymbolColor}{\textbar} \quad <gateCallStatement>
            \\ \textcolor{\grammarSymbolColor}{\textbar} \quad "..."
        \\ \llap{\textcolor{\grammarSymbolColor}{)}\quad}
    
    <\textbf{\textcolor{\grammarRuleDefColor}{annotation}}> ::= <initDefinition>
        \textcolor{\grammarSymbolColor}{\alt} <aodMove>
        \textcolor{\grammarSymbolColor}{\alt} <u3>
        \textcolor{\grammarSymbolColor}{\alt} <rydberg>
        \textcolor{\grammarSymbolColor}{\alt} <annotationKeyword> <remainingLineContent>\textcolor{\grammarSymbolColor}{?}
    
    <\textbf{\textcolor{\grammarRuleDefColor}{initDefinition}}> ::= \textcolor{\grammarTerminalColor}{‘"@init"’} <qubitPositions>

    <\textbf{\textcolor{\grammarRuleDefColor}{qubitPositions}}> ::= \textcolor{\grammarTerminalColor}{‘"["’} <position> \textcolor{\grammarSymbolColor}{(} \textcolor{\grammarTerminalColor}{‘","’} <position> \textcolor{\grammarSymbolColor}{)}\textcolor{\grammarSymbolColor}{*} \textcolor{\grammarTerminalColor}{‘"]"’}

    <\textbf{\textcolor{\grammarRuleDefColor}{position}}> ::= \grammarArg{‘"("’} \syntleft\textcolor{\grammarTypeColor}{float}\syntright  \textcolor{\grammarTerminalColor}{‘","’} \syntleft\textcolor{\grammarTypeColor}{float}\syntright  \textcolor{\grammarTerminalColor}{‘")"’} 
    
    <\textbf{\textcolor{\grammarRuleDefColor}{aodMove}}> ::= \textcolor{\grammarTerminalColor}{‘"@move"’}
        \\ \textcolor{\grammarSymbolColor}{(} \textcolor{\grammarTerminalColor}{‘"row"’} \, \textcolor{\grammarSymbolColor}{|} \, \textcolor{\grammarTerminalColor}{‘"column"’} \textcolor{\grammarSymbolColor}{)} \syntleft\textcolor{\grammarTypeColor}{integer}\syntright  \syntleft\textcolor{\grammarTypeColor}{integer}\syntright

    <\textbf{\textcolor{\grammarRuleDefColor}{u3}}> ::= \textcolor{\grammarTerminalColor}{‘"["’} <rotations> \textcolor{\grammarSymbolColor}{(} \textcolor{\grammarTerminalColor}{‘","’} <rotations> \textcolor{\grammarSymbolColor}{)}\textcolor{\grammarSymbolColor}{*} \textcolor{\grammarTerminalColor}{‘"]"’}
    
    <\textbf{\textcolor{\grammarRuleDefColor}{rotations}}> ::= \grammarArg{'('} \syntleft\textcolor{\grammarTypeColor}{float}\syntright  \syntleft\textcolor{\grammarTypeColor}{float}\syntright  \syntleft\textcolor{\grammarTypeColor}{float}\syntright \grammarArg{')'}
        
    <\textbf{\textcolor{\grammarRuleDefColor}{rydberg}}> ::= \textcolor{\grammarTerminalColor}{‘"@rydberg"’}
    \end{grammar}
    \end{tcolorbox}
    \caption{Abstract grammar for our \qasm in EBNF format. Note that the non-terminals highlighted in purple are renamed from the OpenQASM grammar for simplification purposes. Their definitions, the remaining rules, and the full version of the OpenQASM grammar can be found in OpenQASM specifications~\cite{openqasm2,openqasm3,openqasm3grammar}.
    }
    \Description{}
    \label{fig:grammar}
\end{figure}

%% file: tables/experimental_setup.tex
\begin{figure}[h]
    \centering
    \fontsize{7}{8}\selectfont
    \begin{subfigure}[c]{0.49\textwidth}
        \caption{List of benchmarks used for evaluating \projectname.}
        \centering
        \fontsize{7}{8}\selectfont
        \begin{tabular}{p{0.7\textwidth}|p{0.2\textwidth}}
        \textbf{Algorithm} & \textbf{\# of Qubits} \\ \hline \hline
        BV (Bernstein-Vazirani) \cite{bv_benchmark} & 14, 19 \\
        CAT (Schrödinger Cat Superposition) \cite{monroe_schrodinger_1996} & 22 \\
        QAOA (MaxCut) \cite{wang_quantum_2018} & 14 \\
        DJ (Deutsch-Jozsa) \cite{deutsch_rapid_1997} & 16, 26 \\
        HamSim (Hamiltonian Simulation) \cite{berry_efficient_2007} & 18 \\
        Graph State \cite{hein_entanglement_2006} & 20 \\
        GHZ (Greenberger-Horne-Zeilinger) & 23 \\
        KNN (Quantum k-nearest Neighbors) & 25 \\
        SWP (Swap Test) & 25\\
        WST (W-state) & 24, 27 \\
        Multiply & 13 \\
        \hline
        \end{tabular}
        \label{tab:benchmarks}
    \end{subfigure}
    \begin{subfigure}[r]{0.49\textwidth}
        \caption{QPU hardware parameters.}
        \centering
        \begin{tabular}{p{0.6\linewidth}|p{0.2\linewidth}}
        \fontsize{7}{8}\selectfont
        \textbf{Parameters (adopted from \cite{evered_high-fidelity_2023})} & \textbf{Value} \\ \hline \hline
        Two-qubit gate fidelity  & $0.995$ \\
        Single-qubit gate fidelity  & $0.9991$ \\
        Atom transfer fidelity  & $0.999$ \\
        QPU height  & $155 \mu m$ \\
        QPU width  & $210 \mu m$ \\
        $T_2$  & $1.5 s$ \\
        Atom transfer time  & $17\mu s$ \\
        Atom movement speed  & $0.55\mu\ \mu s$ \\
        Atom acceleration  & $2750m/s^2$ \\
        Single-qubit gate time  & $52\mu s$ \\
        Two-qubit gate time  & $360 ns$ \\    
        \hline
        \end{tabular}
        \label{tab:hardware_parameters}
    \end{subfigure}
    \caption{\textbf{(a)} QPU hardware-model parameters used on \projectname's evaluation. The parameters are based on the published hardware work \cite{evered2023high, Wurtz2023aquila}. \textbf{(b)} List of benchmarks, and respective sizes, used on \projectname's evaluation. These benchmarks were sourced from the QASMBench \cite{li2023qasmbench} open-source benchmark suite.}
    \label{tab:experimental_setup}
    \Description{}
\end{figure}